\documentclass[10pt,a4paper]{article}
 \pdfoutput=1
\usepackage{amsmath,fourier,amssymb,amsthm,graphicx,epstopdf,
color,cases}
\usepackage{chngcntr}
\counterwithout{figure}{section}
\usepackage[T1]{fontenc}
\allowdisplaybreaks
\numberwithin{equation}{section}

 \theoremstyle{plain}
 \newtheorem {hypo}{\bf\hspace{-\parindent}Hypothesis}[section]

 \newtheorem {lemma}[hypo]{Lemma}
 \newtheorem {theo}[hypo]{Theorem}
 \newtheorem {defin}[hypo]{Definition}

 \theoremstyle{remark}
 \newtheorem {rmk}[hypo]{Remark}
  
 \newcommand{\pf}{\begin{bpf}}

 \newcommand{\pfms}{\begin{bpfms}}
 \newcommand{\epf}{\end{bpf}\hfill$\square$\vspace{0.1cm}}
 \newcommand{\epfms}{\end{bpfms}\hfill$\square$\\ }
 \newcommand\ben{\begin{equation*}}
 \newcommand\ebn{\end{equation*}}
 \newcommand\beq{\begin{equation}}
 \newcommand\eeq{\end{equation}}
 \newcommand\ds{\displaystyle}
  \newcommand\lb{\left(}
  \newcommand\rb{\right)} 
  
   \newcommand\Cb{\mathbb{C}} 
   \newcommand\Zb{\mathbb{Z}}
   \newcommand\Pb{\mathbb{P}}

 \def\mc{\mathcal}
\newcommand{\eq}[1]{\begin{equation}\begin{gathered}
#1\end{gathered}\end{equation}}
\def\Sum{\sum\limits}

\def\d{\partial}

\begin{document}
\LARGE
\noindent
\textbf{Pure $\mathrm{SU}\lb 2\rb$ gauge theory partition function \\ and generalized Bessel kernel}
\normalsize
 \vspace{1cm}\\
 \noindent\textit{
 P. Gavrylenko$\,^{a,b,c,}$\footnote{pasha145@gmail.com}, 
 O. Lisovyy$\,^{d,}$\footnote{lisovyi@lmpt.univ-tours.fr}}
 \vspace{0.2cm}\\
 $^a$ Center for Advanced Studies, Skolkovo Institute of Science and Technology, 143026 Moscow, Russia
  \vspace{0.1cm}\\
 $^b$ National Research University Higher School of Economics, Department of
 Mathematics and International Laboratory of Representation Theory and Mathematical
 Physics, 119048 Moscow, Russia\vspace{0.1cm}\\
 $^c$ Bogolyubov Institute for Theoretical Physics,  03680 Kyiv, Ukraine
 \vspace{0.1cm}\\
 $^d$ Laboratoire de Math\'ematiques et Physique Th\'eorique CNRS/UMR 7350,  Universit\'e de Tours, Parc de Grandmont,
  37200 Tours, France
  
  \begin{abstract}
  We show that the dual partition function of the pure $\mathcal N=2$ $\mathrm{SU}\lb 2\rb$ gauge theory in the self-dual $\Omega$-background (a) is given by
  Fredholm determinant of a generalized Bessel kernel and (b) coincides with the tau function associated to the general solution of the Painlevé III equation of type $D_8$ (radial sine-Gordon equation). In particular, the principal minor expansion of the Fredholm determinant yields Nekrasov combinatorial sums over pairs of Young diagrams.
  \end{abstract}
  
 \section{Introduction} 
 
 The study of quantitative aspects of the isomonodromy/CFT correspondence \cite{SMJ,Knizhnik,Moore,Teschner} has been initiated in the work  \cite{GIL12}, where the general tau function
 of the sixth Painlev\'e equation was conjectured to coincide with the Fourier tranform of the 4-point $c=1$ Virasoro conformal block
 \eq{\label{pvicft}
 \tau_{\mathrm{VI}}\bigl( t\,|\,\sigma,\eta,\vec\theta\bigr)=\Sum_{n\in\mathbb{Z}}e^{2\pi in\eta}\vcenter{\hbox{\includegraphics[height=1.5cm]{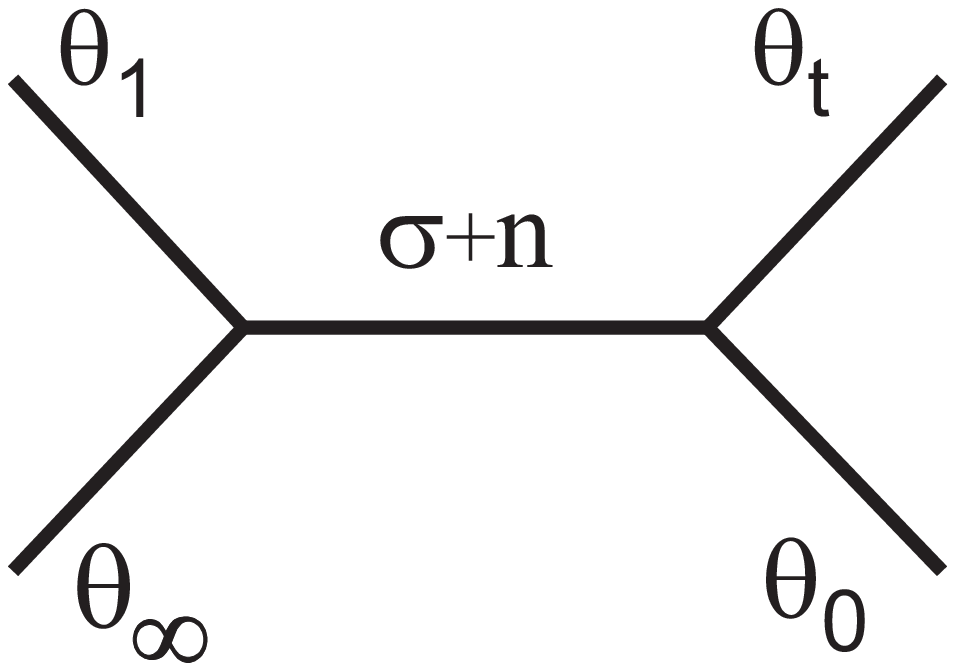}}}
 (t). }
 This proposal was later proved in \cite{ILTe,BSh1} by CFT methods. The parameters $\vec\theta=\lb\theta_0,\theta_t,\theta_1,
 \theta_\infty\rb$  represent local monodromy exponents on the Painlevé side, and are related to external conformal dimensions of primaries in the conformal block by $\Delta_{\nu}=\theta_{\nu}^2$. The intermediate dimension is $\Delta=(\sigma+n)^2$. 
 
 As is well-known,
 the AGT correspondence \cite{AGT} relates Virasoro 4-point conformal  blocks to partition functions of the $\mc N=2$ supersymmetric 4D gauge
 theories with the gauge group $\mathrm{SU}\lb 2\rb$ and $N_f=4$ matter multiplets, regularized by an appropriate deformation (the $\Omega$-background) with two parameters
 $\epsilon_1, \epsilon_2$. The $c=1$ case corresponds to the self-dual $\Omega$-background ($\epsilon_1+\epsilon_2=0$). Expanding conformal blocks around $t=0$ corresponds to the  weak coupling expansion in the gauge theory, explicitly  computed in \cite{Nekrasov}.
 
 The Painlev\'e VI is the most general equation in the Painlevé family. All the others can be obtained from it by appropriate 
 degeneration limits. In \cite{GIL13}, some of these limits have been  computed at the level of solutions. This produces explicit formulas for Painlevé V and all three types ($D_6$, $D_7$ and $D_8$) of Painlevé III functions  in the form of power series. From the gauge theory point of view, such degenerations correspond to decoupling of the massive fields, which means that Painlev\'e V and III's are related to $N_f<4$ gauge theories, and explicit formulas for the tau functions are known in their weak coupling regions.
 On the CFT side, these cases are related to conformal blocks involving Whittaker vectors \cite{Gaiotto,Bonelli,GT}. In contrast to the $N_f=4$ case, there are interesting situations for $N_f<4$ where explicit (asymptotic) series representations of solutions
 are not known: they correspond to strong coupling regions on the gauge theory side, and to conformal blocks with irregular
 vertex operators in the CFT framework. 
 The present work is concerned with the most degenerate
 case of Painlevé III equation of type $D_8$ corresponding to the pure gauge theory.

 It is interesting to note that an avatar of the
 Painlevé III ($D_8$) tau function was already studied by Nekrasov and Okounkov in \cite{NO}, although  at the time the relevant object had not yet been related to isomonodromy nor to CFT. The equation of interest is of 2nd order and contains no parameters;  its tau function is given~by
 \eq{\label{piiicft}
 \tau_{\mathrm{III}}\lb t\,|\,\sigma,\eta\rb=\Sum_{n\in\mathbb Z} e^{4\pi i n\eta}  \mc Z_{\mathrm{SU}\lb 2\rb}\lb t\,|\,\sigma+n\rb,
 }
 where $(\sigma,\eta)$ represent the initial data.
 The right side of (\ref{piiicft}) was dubbed in \cite{NO} the dual partition function of the pure gauge theory. The first reason to consider it was purely technical: it is 
 convenient to  introduce a Lagrange multiplier to control (in the non-$\epsilon$-deformed limit) the value of $\sigma$, the vacuum expectation value 
 of the scalar field. A second reason is the existence of a fermionic representation for the dual partition function,  presented in \cite{NO} in the special case $\tau_{\mathrm{III}}\lb t\,|\,\frac14,\eta\rb$. Setting in addition $\eta=0$ or $\eta=\frac14$,  we obtain elementary solutions of PIII:
 \beq
 \tau_{\mathrm{III}}\lb t\,\Bigl|\,\text{\footnotesize$\frac14$},
 \text{\footnotesize$\frac{1\pm1}{8}$}\rb=t^{\frac1{16}}e^{\pm 4\sqrt t}.
 \eeq
They are related to twisted representations in the intermediate channel \cite{Zam,Apikyan} generated by the realization of the Virasoro algebra in terms of one Ramond boson \cite{BSh2}.
 
 In order to get a physically interesting result, namely the partition function without $\epsilon$-deformation, the dual partition function should be considered in the
 limit $\eta\to i\infty$. In this case the sum can be computed in a saddle-point approximation\footnote{This is the original proposal from \cite[Eq. (5.5)]{NO}. The actual answer for the dual partition function also contains non-perturbative corrections (in $\epsilon$) of crucial importance  which we are going to study in a future work.}. Different quantities scale as follows:
 \beq
 \begin{gathered}
 \eta=\epsilon^{-1}\tilde\eta,\qquad\sigma=\epsilon^{-1}\tilde\sigma,\qquad t=\epsilon^{-4}\tilde t,\\
 \mc Z_{\mathrm{SU}\lb 2\rb}\left(\epsilon^{-4}\tilde t\,|\,\epsilon^{-1}\tilde\sigma\right)\sim
 \exp\left\{\epsilon^{-2}\mc F_0(\tilde t\,|\,\tilde\sigma)+\mc F_1(\tilde t\,|\,\tilde\sigma)+\ldots\right\},
 \end{gathered}
 \eeq
 which means that the saddle point is defined by the equation
 $\d_{\tilde\sigma}\mc F_0\lb\tilde\sigma\,|\,\tilde t\rb=-4\pi i\tilde\eta$.
  One of the main results of \cite{NO} is the statement that the Seiberg-Witten prepotential \cite{SW} --- the function encoding the low-energy behaviour of the $\mc N=2$ pure $\mathrm{SU}\lb 2\rb$ gauge theory --- coincides with $\mc F_0\lb\tilde\sigma\,|\,\tilde t\rb$, which confirms the Seiberg-Witten solution at the microscopic level.

 A related procedure was used in \cite{BLMST} to identify the Painlevé I--V tau functions also with the dual partition functions of \textit{strongly} coupled gauge theories, including the Argyres-Douglas theories of type $H_{0}$, $H_1$ and~$H_2$. Specifically, it has been checked that the long-distance (irregular type) tau function expansions  match various magnetic and dyonic strong coupling expansions on the gauge side. A CFT counterpart of this correspondence has been suggested in \cite{Nagoya,Nagoya2}, where some of the long-distance asymptotic  series for Painlevé~V and IV were conjecturally related to Fourier transforms of conformal blocks 
 with irregular vertex operators.
 
 In a recent paper \cite{GL16}, we have developed a method of representing the isomonodromic tau functions of Fuchsian systems as block Fredholm determinants. The construction is based on the Riemann-Hilbert approach. The main input is given by  monodromy of a connection $\partial_z- A\lb z\rb$ with simple poles together with a pants decomposition of the appropriate punctured Riemann sphere. The relevant integral operators act on vector-valued functions defined on a collection of circles (internal boundary components of pants). Their kernels are expressed in terms of solutions of Fuchsian systems associated to different pairs of pants and having only 3 regular singular points. In rank 2, where the isomonodromy equations are equivalent to the Garnier system containing Painlevé VI as the simplest case, Fredholm determinant representations become completely explicit
 as the kernels have hypergeometric expressions. Furthermore, the principal minor expansion of the determinant written in the Fourier basis coincides with  the combinatorial evaluation \cite{Nekrasov} of the dual partition function of the 4D $\mathcal N=2$  linear quiver $\mathrm{U}\lb 2\rb$ gauge theory. This yields in particular a rigorous proof of the series representation of the Painlevé~VI tau function, which bypasses the use of the AGT correspondence and does not rely on CFT arguments such as crossing symmetry, null vector decoupling equations,~etc.
 
 While it is in principle clear that the approach of \cite{GL16} may be extended to at least some classes of irregular isomonodromic systems, its practical implementation  within the Riemann-Hilbert framework  is not obvious. Our main goal in this paper is to work out the details for Painlevé III ($D_8$) equation which exhibits most of the subtleties of the irregular case and at the same time keeps  the notational fuss to a minimum. We hope that the Fredholm determinant representation of  $\tau_{\mathrm{III}}\lb t\,|\,\sigma,\eta\rb$ obtained here, besides producing a combinatorial series at weak coupling, may also turn out to be useful for the analysis of the strongly coupled regime. Let us mention that a different (?) Fredholm determinant representation for the special tau function $\tau_{\mathrm{III}}\lb t\,|\,\sigma,0\rb$ has recently appeared in the proof \cite{BGT1} of a 4D version of the conjecture of \cite{GHM} relating topological strings and spectral theory (see also \cite{BGT2} for higher-rank generalizations). Our results could also provide some insight in this context.  
 
\begin{figure}[h!]
                 \centering
                 \includegraphics[height=2.2cm]{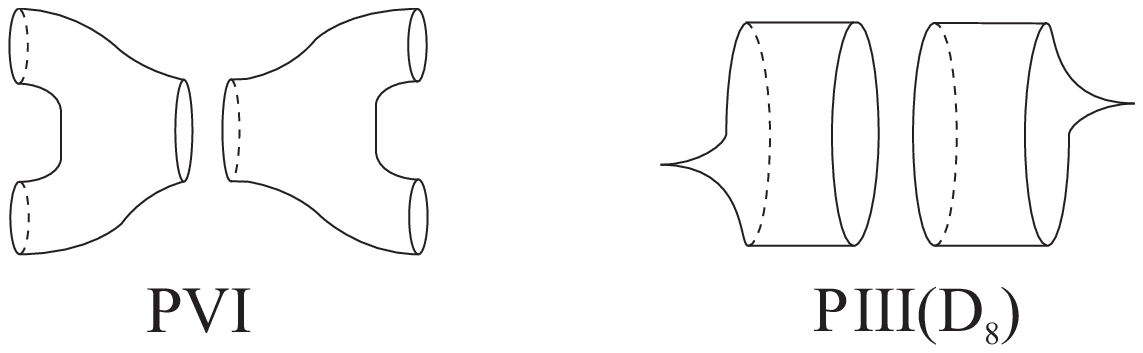}
                 \caption{Pants decomposition for Painlevé VI and Painlevé III ($D_8$).\label{p3pants}}
                 \end{figure}

 A useful guideline for our work is provided by the geometric Painlevé confluence diagram proposed in \cite{CM,CMR}. In this picture, the monodromy manifolds of different Painlevé equations are interpreted as moduli spaces of Riemann spheres with cusped boundaries. One is then tempted to replace the usual decomposition of the Painlevé VI four-holed sphere into two pairs of pants by cutting the Pain\-levé~III~($D_8$) decorated cylinder into two, each of them having one regular and one 1-cusped puncture, see Fig.~\ref{p3pants}.

 \noindent Furthermore, the number of cusps at a particular hole was heuristically related \cite[Appendix A]{CMR} to the number of Stokes rays at the corresponding irregular singular point, and to the pole order of the quadratic differential $\operatorname{det}A\lb z\rb\,dz^2$. As we will see, the former interpretation turns out to be the most adapted to our purposes, cf e.g. the Riemann-Hilbert contour in Fig.~\ref{figgammahat}.

 The paper is organized as follows. In Section~\ref{sec_iso}, we introduce an irregular linear system leading to Pain\-levé~III ($D_8$), describe its generalized monodromy, and explain the ``decorated pants decomposition'' of the associated Riemann-Hilbert problem. In Section~\ref{sec_fred}, it is shown that the PIII ($D_8$) tau function admits a Fredholm determinant representation with a generalized Bessel kernel, the main result being 
 Theorem~\ref{theoFr}. Section~4 is devoted to derivation of the series over pairs of Maya/Young diagrams and its identification with the dual partition function of the pure gauge theory (Theorems~\ref{thMaya} and~\ref{thYoung}). \vspace{0.1cm}
 
   \noindent
   { \small \textbf{Acknowledgements}.  We would like to thank M. Bershtein, N. Iorgov,  and A. Marshakov for useful discussions. The present work was supported by the CNRS/PICS project ``Isomonodromic deformations and conformal field theory''. P.~G. was partially supported by the RSF grant No. 16-11-10160 (results of section 4). He is also a Young Russian Mathematics award winner and would like to thank its sponsors and jury. P.~G. would also like to thank the KdV Institute of the University of Amsterdam, where a part of this work was done, and especially G.~Helminck, for warm hospitality.}

 \section{Isomonodromy and  Riemann-Hilbert setup\label{sec_iso}}

 \subsection{Associated irregular system}
 Our starting point is a system of linear differential equations 
 \begin{subequations}\label{LSsub}
 \beq\label{LS}
 \partial_z Y= A\lb z\rb  Y,
 \eeq
 where $A\lb z\rb$ is a given $N\times N$ matrix with rational dependence on $z$. The fundamental matrix solution $Y\lb z\rb$ in general has branched singularities at the poles of the 1-form $A\lb z\rb dz$ on the Riemann sphere $\Pb^1$.
 It involves no loss of generality to assume that $\operatorname{Tr} A\lb z\rb=0$; otherwise it suffices to transform $Y\mapsto f Y$ with a suitably adjusted scalar factor.
 
 We are going to study a special class of such linear systems in rank $N=2$ characterized by the number of singularities and their type. Specifically, assume that there are only two irregular singular points (e.g. $0$ and $\infty$) of Poincaré rank $\frac12$. By this we mean that 
 \ben
 A\lb z\rb=A_{-2}z^{-2}+A_{-1}z^{-1}+A_0,\qquad 
 A_k\in\operatorname{Mat}_{2\times 2}\lb \Cb\rb,\quad 
 \operatorname{Tr}A_k=0,
 \ebn
 with non-diagonalizable $A_0$ and $A_{-2}$. Using constant gauge
 transformations $Y\mapsto G Y$, $A\lb z\rb\mapsto
 G A\lb z\rb G^{-1}$ and rescaling $z\mapsto\lambda z$ if necessary, it may be further assumed that either
(i) $A_0=\sigma_+$, $A_{-2}=\sigma_-$ or (ii) $A_0=A_{-2}=\sigma_+$, where
 \ben
 \sigma_+=\lb \begin{array}{cc} 0 & 1 \\ 0 & 0 \end{array}\rb,\qquad
 \sigma_-=\lb \begin{array}{cc} 0 & 0 \\ 1 & 0 \end{array}\rb.
 \ebn
 In the case (ii), the remaining freedom of conjugation by upper triangular matrices with unit diagonals leaves only two nontrivial parameters in $A_{-1}$. The corresponding linear system does not admit
 isomonodromic deformations and reduces to a special case of doubly-cofluent Heun equation. We will therefore focus on the case~(i) and, after suitable rescalings, parameterize
 $A\lb z\rb$ as
 \beq\label{ALS}
 A\lb z\rb=q\sigma_- z^{-2}+ q^{-1}\lb\begin{array}{cc} -p & t \\ -q & p
 \end{array}\rb z^{-1} -\sigma_+.
 \eeq
 \end{subequations}
 The system (\ref{LS}) with $A\lb z\rb$ given by (\ref{ALS}) is the linear problem associated to Painlevé III ($D_8$) equation. Among 3 parameters $p$, $q$ and $t$, the latter plays the role of time in the associated isomonodromic problem, and the former two are coordinates on the PIII ($D_8$) phase space.
 
 The system (\ref{LSsub}) can be put to a more convenient form using  non-constant gauge transformation. Let us define a
 new matrix $\tilde Y\lb \xi\rb$ by
 \ben
  \tilde Y\lb \xi\rb={G\lb \xi\rb}^{-1}  Y\lb \xi^2\rb,\qquad G\lb \xi\rb=\frac1{i\sqrt2}\lb \begin{array}{cc} 
 \xi^{\frac12} & \xi^{\frac12} \\ \xi^{-\frac12} & -\xi^{-\frac12}
 \end{array}\rb.
 \ebn
 It solves the linear system
 \begin{subequations}\label{LStrans2}
 \beq\label{LStrans}
 \partial_{\xi} \tilde Y = \tilde A \lb \xi\rb \tilde Y,
 \eeq
 with $ \tilde A \lb \xi\rb = 2\xi {G\lb \xi\rb}^{-1}  A\lb \xi^2\rb G\lb \xi\rb -{G\lb \xi\rb}^{-1}G'\lb \xi\rb$. Computing  the latter matrix explicitly, one may see that the system (\ref{LStrans}) also has irregular singularities at $0$ and $\infty$:
 \beq\label{LStransb}
 \begin{gathered}
 \tilde{A}\lb \xi\rb= \tilde A_{-2}\xi^{-2}+ \tilde A_{-1}\xi^{-1}+ \tilde A_{0},\\
 \tilde A_{-2} =\lb q + \frac tq  \rb\sigma_z+\lb q- \frac  tq  \rb i\sigma_y,\qquad
 \tilde A_{-1}=-\lb\frac {2p}q+\frac12\rb\sigma_x,\qquad
 \tilde A_0=-2\sigma_z,
 \end{gathered}
 \eeq
 where $\sigma_{x,y,z}$ denote the Pauli matrices. The above is by no means a generic form of $2\times 2$ systems with 2 irregular singular points of Poincaré rank 1; one of the properties that singles out the class described by (\ref{LStransb}) is a discrete $\mathbb Z_2$-symmetry $ \tilde A\lb -\xi\rb=-\sigma_x \tilde A\lb \xi\rb\sigma_x$.

 \end{subequations}

  \subsection{Monodromy\label{subsecMonodromy}}
   The fact that the transformed coefficients $\tilde A_{-2}$ and $\tilde A_0$ are  diagonalisable, in contrast to their counterparts in (\ref{ALS}), allows to write formal fundamental solutions of (\ref{LStrans}) at $0$ and $\infty$ in the standard form,
   \begin{subequations}\label{formalsols}
  \begin{align}
  \label{formalsol0}
  &  \tilde Y^{(0)}_{\text{form}}\lb \xi\rb=
   \lb -\frac{q}{\sqrt t}\rb^{-\frac{\sigma_x}2}\left[\mathbb 1+\sum_{k=1}^{\infty}y^{(0)}_k \xi^{k}\right]
   e^{2\sigma_z\sqrt t\,\xi^{-1}},\qquad\xi\to0,\\
   \label{formalsol8}
  & \tilde Y^{(\infty)}_{\text{form}}\lb \xi\rb=
  \left[\mathbb 1+\sum_{k=1}^{\infty}y^{(\infty)}_k \xi^{-k}\right]
  e^{-2\sigma_z\xi},\qquad\qquad\qquad\quad\;\;\xi\to\infty.
  \end{align}
  \end{subequations}
  The $\mathbb Z_2$-symmetry of $\tilde A\lb \xi\rb$ implies that formal solutions satisfy $\tilde Y^{(\nu)}_{\text{form}}\lb -\xi\rb
  =\sigma_x \tilde Y^{(\nu)}_{\text{form}}\lb \xi\rb\sigma_x$.
    The expansion coefficients $y^{(\nu)}_k$ can be computed in a straightforwad way to any finite order using (\ref{LStrans2}). In what follows, the only explicit expression we need concerns the first such coefficient in (\ref{formalsol0}), namely,
  \beq\label{expcoef1}
  y^{(0)}_1=-\frac{1}{\sqrt t}\left[\lb\frac{p^2}{q^2}+\frac{p}{2q}-q-\frac tq+\frac{1}{16}\rb\sigma_z+\left(\frac{p}{2q}+\frac18\right)i\sigma_y\right].
  \eeq
  
  The actual solutions of the unfolded system (\ref{LStrans2}) can only be asymptotic to $\tilde Y^{(\nu)}_{\text{form}}\lb \xi\rb$ inside the Stokes sectors $\tilde {\mathcal S}_{k}^{(\nu)}$ ($k=1,2,3$) defined~by
  \begin{align*}
  \tilde{\mathcal S}_k^{(0)}&\,=\left\{\xi\in\Cb\,\bigl|\, \frac{\arg t-3\pi}{2}+k\pi<\arg\xi<\frac{\arg t+\pi}{2}+k\pi,\;|\xi|<R\right\},\\
  \tilde{\mathcal S}_k^{(\infty)}&\,=\left\{\xi\in\Cb\,\bigl|\, -\frac{3\pi}{2}+k\pi<\arg\xi<\frac{\pi}{2}+k\pi,\;|\xi|>R\right\}.
  \end{align*}
  Furthermore, the requirement that $\tilde Y^{(\nu)}_{k}\simeq  \tilde Y^{(\nu)}_{\text{form}}$ inside $\mathcal S_{k}^{(\nu)}$ as $\xi\to\nu$ fixes the  solutions
  $\tilde Y^{(\nu)}_{k}$  uniquely. As is well-known, such canonical solutions associated to the same point are related by constant (i.e. independent of~$\xi$) Stokes matrices 
  \beq\label{StokesM}
  S^{(\nu)}_{k}={\tilde Y^{(\nu)}_{k}\lb\xi\rb}^{-1}
  \tilde  Y^{(\nu)}_{k+1}\lb\xi\rb,\qquad \nu=0,\infty,\qquad k=1,2.
  \eeq
  Constant connection matrix $E={\tilde Y^{(0)}_{1}\lb\xi\rb}^{-1}
     \tilde Y^{(\infty)}_{1}\lb\xi\rb$ relates the canonical solutions at $0$ and $\infty$. Stokes and connection matrices describe global asymptotic behavior of solutions of  (\ref{LSsub}), (\ref{LStrans2}) and constitute the relevant set of generalized monodromy data.
    
    The discrete symmetry  mentioned above and interlacing structure of dominant/recessive solutions imply that Stokes matrices can be written as
    \begin{subequations}\label{symcons}
    \begin{align}
    S^{(0)}_{1}=\sigma_x S^{(0)}_{2}\sigma_x=\lb
    \begin{array}{cc}
    1 & \alpha \\ 0 & 1
    \end{array}\rb,\qquad
    S^{(\infty)}_{1}=\sigma_x S^{(\infty)}_{2}\sigma_x=
    \lb
        \begin{array}{cc}
        1 & 0 \\ \beta & 1
        \end{array}\rb.
    \end{align}
    Similar constraints are also valid for the connection matrix,
    \beq
    \sigma_x E\sigma_x={S_1^{(0)}}^{-1}E\,S^{(\infty)}_1={S_2^{(0)}}E\,{S^{(\infty)}_2}^{-1}.
    \eeq
    \end{subequations}
    The relations (\ref{symcons}) imply that, in general, the monodromy data $\bigl\{S^{(\nu)}_k\bigr\}$, $E$ can be parameterized by a pair of complex parameters ${\lb \sigma,\eta\rb}$ in the following way:
    \beq\label{monodrparam}
    \begin{aligned}
    &E=\frac{1}{\sin2\pi\sigma}\lb\begin{array}{cc}
    \sin2\pi\eta & -i\sin2\pi\lb \eta+\sigma\rb \\
    i\sin2\pi\lb\eta-\sigma\rb & \sin2\pi\eta
    \end{array}\rb,\qquad \sigma\notin\mathbb Z/2,\\
    &S_{1}^{(0)}={S_{2}^{(0)}}^T={S_{1}^{(\infty)}}^T
    =S_{2}^{(\infty)}\equiv S=\lb\begin{array}{cc}
    1 & -2i\cos2\pi\sigma \\ 0 & 1
    \end{array}\rb.
    \end{aligned}
    \eeq
    It can be furthermore assumed that $\sigma$ and $\eta$ belong to the strips $-\frac12\leq \Re\sigma\leq0$ and
    $-\frac12<\Re\eta\leq\frac12$. Note that the counterclockwise monodromy matrix $\tilde M_0$ of $\tilde Y^{(0)}_1\lb z\rb$ around $0$ can be expressed as
    \begin{align}
    \label{monM0}
    &\tilde M_0^{-1}=SS^T=ES^TSE^{-1}.
    \end{align}
    
    Let us finally comment on how to recover monodromy of the initial system (\ref{LSsub}) from the Stokes data of the unfolded equation (\ref{LStrans2}). Introduce the solutions $Y^{(\nu)}\lb z\rb=G\lb \sqrt z\rb\tilde Y_1^{(\nu)}\lb \sqrt z\rb$, uniquely defined by their asymptotic behavior
    \ben
    Y^{(\nu)}\lb z\rb\simeq G\lb \sqrt z\rb\tilde Y_{\text{form}}^{(\nu)}\lb \sqrt z\rb,
    \ebn 
    as $z\to \nu$ inside the sectors 
   $\arg t-\pi<\arg z <\arg t+3\pi$ (for $\nu=0$) and
    $-\pi<\arg z <3\pi$ (for $\nu=\infty$). The monodromy matrix in $Y^{(0)}\lb z e^{2\pi i}\rb=Y^{(0)}\lb z\rb M_0$ can be computed using the Stokes matrix connecting unfolded solutions $\tilde Y_{1,2}^{(0)}\lb\xi\rb$ together with the symmetry properties
    $G\lb \xi e^{i\pi}\rb=iG\lb\xi\rb\sigma_x$ and  $\sigma_x \tilde Y_{\mathrm{form}}^{(0)}\lb -\xi\rb\sigma_x=\tilde Y_{\mathrm{form}}^{(0)}\lb\xi\rb$. The result reads
        \begin{subequations}
    \begin{align}
    \label{eigenM0}
    &M_0=i\sigma_x S^{-1}=\lb\begin{array}{cc}
    0 & i \\ i & -2\cos2\pi\sigma
    \end{array}\rb = U^{-1}e^{2\pi i\mathfrak S} U,\\
    \label{matrU}
        & \mathfrak S=\lb\sigma+\frac12\rb\sigma_z,\qquad
        U=\frac{1}{\sqrt{2\sin2
        \pi\sigma}}
        \lb\begin{array}{cc}e^{-i\pi \lb\sigma +\frac14\rb} & e^{i\pi  \lb\sigma+\frac14\rb} \\ e^{i\pi\lb\sigma+\frac14\rb} & -e^{-i\pi\lb\sigma+\frac14\rb}\end{array}\rb.
        \end{align}
        \end{subequations}
    As expected, the monodromy matrices of $Y^{(0)}$ and $\tilde Y^{(0)}_1$ are related by $\tilde M_0=-M_0^2$. In the same way, the monodromy of $Y^{(\infty)}\lb z\rb$ around $0$ is given by $E^{-1}M_0E=\sigma_x M_0\sigma_x$.

    \subsection{Deformation equations and tau function}
  The usual construction of isomonodromic family of systems (\ref{LStrans2}) involves varying the ``time'' parameter $t$ appearing in the exponentials in (\ref{formalsols}), while keeping the data  $\bigl\{S^{(\nu)}_k\bigr\}$, $E$ fixed. The latter requirement implies that the matrix $\partial_t \tilde Y\cdot{ \tilde Y}^{-1}$ is meromorphic on $\Pb^1$ with poles only possible at $0$ and
  $\infty$. Analyzing the local behavior of this quantity with the help of expansions of formal solutions and recasting the result in terms of $Y\lb z\rb$, one finds that
  \ben
  \partial_t Y= B\lb z\rb  Y,\qquad
   B\lb z\rb=\lb\begin{array}{cc}
   0 & -q^{-1} \\
   -\frac{q}{tz} & 0\end{array}\rb.
  \ebn
  The compatibility of this isomonodromy constraint with the system (\ref{LSsub}) yields the zero-curvature condition
  $\partial_t A-\partial_{z} B+[ A, B]=0$,
  which is equivalent to a pair of scalar equations
 \beq\label{hamform}
 \begin{cases} tq_t=2p+q,\\
 tp_t=\ds\frac{2p^2}{q}+p+q^2-t,
 \end{cases}
 \eeq
 or, equivalently, to a single 2nd order ODE:
 \beq\label{PIIID8}
 q_{tt}=\frac{q_t^2}{q}-\frac{q_t}{t}+\frac{2q^2}{t^2}-\frac2t.
 \eeq
 This is the most degenerate Painlevé III equation (of type $D_8$).
 In applications, it usually appears in the form of the radial sine-Gordon equation
 \beq\label{rSG}
 u_{rr}+\frac{u_r}{r}+\sin u=0,
 \eeq
 which is obtained from (\ref{PIIID8}) after the change of variables $q\lb 2^{-12}r^4\rb = -2^{-6}r^2 e^{iu\lb r\rb}$. The isomonodromic provenance of these equations implies that the quantities $\lb\sigma,\eta\rb$ introduced above to parameterize the Stokes data provide a pair of conserved quantities for (\ref{PIIID8}) and
 (\ref{rSG}).
 
 Let us define the tau function $\tau_{\text{III}}\lb t\rb$ of PIII ($D_8$) by the logarithmic derivative
 \beq\label{deftau}
 \zeta\lb t\rb=t\partial_t\ln\tau_{\text{III}}=\frac{\lb tq_t-q\rb^2}{\;4q^2}-q-\frac tq.
 \eeq
 Conversely, one can express $q=-t\zeta'$.
 The function $\zeta\lb t\rb$ essentially coincides with the time-dependent Hamiltonian of PIII ($D_8$) and satisfies the equation 
 \beq\label{sigmapiii}
 \lb t\zeta''\rb^2=4\lb\zeta'\rb^2\lb\zeta-t\zeta'\rb-4\zeta'.
 \eeq
 The tau function plays a crucial role in the rest of this note.
 We are going to express it in terms of monodromy, thereby providing explicit formulas for the general solution of Painlevé III ($D_8$).
  
 \subsection{Riemann-Hilbert problem} 
\begin{figure}[h!]
                \centering
                \includegraphics[height=6cm]{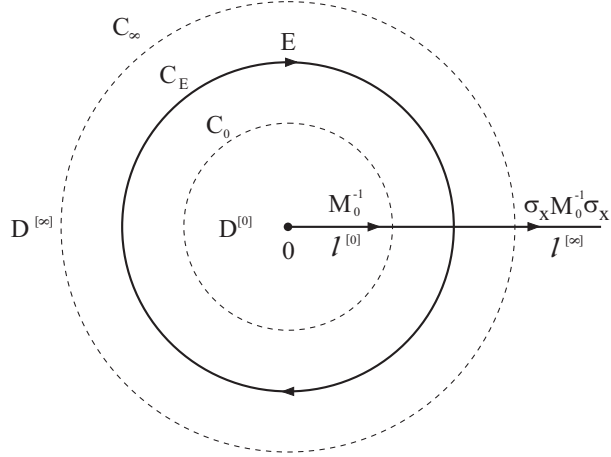}
                \caption{\label{RHPini} Contour $\Gamma$ of the PIII ($D_8$) Riemann-Hilbert problem.\label{figgamma}}
                \end{figure}

 It is convenient to replace the linear system (\ref{LStrans2}) by an equivalent Riemann-Hilbert problem (RHP). It will be defined by a pair $\lb\Gamma,J\rb$ where $\Gamma$ is an oriented contour on $\Pb^1$ and
 $J:\Gamma\to\mathrm{SL}\lb 2,\Cb\rb$ is a jump matrix. The relevant contour $\Gamma=\ell^{[0]}\cup \ell^{[\infty]}\cup C_E$ is represented by solid lines in Fig.~\ref{figgamma}  where it is assumed for simplicity that $\arg t=0$.  The segments $\ell^{[0]}$ and $\ell^{[\infty]}$ correspond to portions of anti-Stokes rays at $0$ and $\infty$. They are close relatives of the cusps of the Chekhov-Mazzocco-Rubtsov geometric confluence diagram \cite{CM,CMR} in the Riemann-Hilbert setting.

 The Riemann sphere is decomposed by $\Gamma$ into 2 connected open domains $D^{[0]}$ and $D^{[\infty]}$. The relevant RHP is to find a $2\times 2$ matrix $\Psi\lb z\rb$ holomorphic and invertible inside each of these domains such that
 \begin{enumerate}
 \item[(i)] its boundary values on the positive and negative side of $\Gamma$ satisfy $ \Psi_+=\Psi_-J$, where the piecewise constant jump matrix is given by
 \ben
 J\bigl|_{\ell^{[0]}}=M_0^{-1},\qquad
 J\bigl|_{\ell^{[\infty]}}=\sigma_x M_0^{-1}\sigma_x,\qquad
 J\bigl|_{C_E}=E.
 \ebn
 \item[(ii)] at $0$ and $\infty$, the function $\Psi\lb z\rb$ behaves as
 \begin{align*}
 \Psi\lb z\rb\simeq G\lb\sqrt{z}\rb\begin{cases}
 Q\bigl[\mathbb 1+O\lb \sqrt z\rb\bigr]\,
 e^{2\sigma_z\sqrt{t/z}},\quad &z\to0,\vspace{0.1cm}\\
 \left[\mathbb 1+O\lb \frac{1}{\sqrt z}\rb\right]\,
  e^{-2\sigma_z\sqrt z},\quad & z\to\infty,
 \end{cases}
 \end{align*} 
  where $\arg z\in ]-\pi,\pi[$ and $Q$ is a constant invertible matrix such that $[Q,\sigma_x]=0$, cf (\ref{formalsols}).
  \end{enumerate}
 The unique solution of this RHP is related to the fundamental solution of the irregular system (\ref{LSsub}) by
 \ben
  Y^{(0)}\lb z\rb=\begin{cases}
 \Psi\lb z\rb,\qquad & z\in D^{[0]},\\
 \Psi\lb z\rb E^{-1},\qquad & z\in D^{[\infty]}.
 \end{cases}
 \ebn
 
\begin{figure}[h!]
                 \centering
                 \includegraphics[height=5cm]{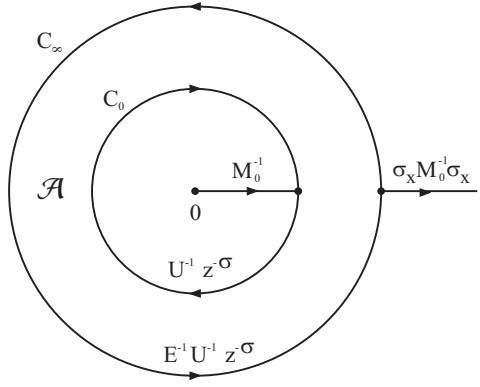}
                 \caption{\label{RHPhat} Contour $\hat\Gamma$ and associated jumps. \label{figgammahat}}
                 \end{figure}

 Let $\mathcal A$ be the open annulus bounded by two circles $C_0$ and $C_{\infty}$ as shown in Fig.~\ref{figgamma}. Consider a piecewise analytic function $\hat\Psi\lb z\rb$ defined by
 \beq\label{Psihat}
 \hat\Psi\lb z\rb=\begin{cases}
 {\Psi}\lb z\rb,\qquad & z\notin\bar{\mathcal A},\\
  Y^{(0)}\lb z\rb U^{-1} z^{-\mathfrak S}, 
 \qquad & z\in\mathcal A,
 \end{cases}
 \eeq
 where $\mathfrak S=\lb \sigma+\frac12\rb\sigma_z$ and $U$ are defined by (\ref{matrU}).
 This new function solves a Riemann-Hilbert problem defined by the pair $\lb\hat\Gamma,\hat J\rb$, where the contour $\hat \Gamma$ and the relevant jump matrices are represented in Fig.~\ref{figgammahat}. The transformed RHP is of course equivalent to the initial one. The function $\hat\Psi\lb z\rb$ has been designed so that it has no jumps inside $\mathcal A$ and coincides with $\Psi\lb z\rb$ inside $C_0$ and outside $C_{\infty}$. Cancellation of the jumps inside $\mathcal A$ can only be done at the expense of introducing new jumps on the circles $C_{0}$ and $C_{\infty}$; as we will see in a moment, their choice above models regular singularities at $\infty$ and $0$, respectively.
 
 There is a natural decomposition $\hat\Gamma=
 \hat\Gamma^{[0]}\cup\hat\Gamma^{[\infty]}$, where $\hat\Gamma^{[0]}$ (and $\hat\Gamma^{[\infty]}$) consist of  $C_{0}$ (resp. $C_{\infty}$) and the part of the positive real axis contained
 inside $C_0$ (resp. outside $C_{\infty}$). Denoting
 \ben
 \hat J^{[0]}=\hat{J}\Bigl|_{\hat\Gamma^{[0]}},\qquad
 \hat J^{[\infty]}=\hat{J}\Bigl|_{\hat\Gamma^{[\infty]}},
 \ebn 
 we can assign to the original RHP two simpler RHPs for functions
 $\hat\Psi^{[0]}\lb z\rb$ and $\hat\Psi^{[\infty]}\lb z\rb$ defined by the pairs
 $\lb \hat\Gamma^{[0]},\hat J^{[0]}\rb$ and  $\lb \hat\Gamma^{[\infty]},\hat J^{[\infty]}\rb$. The latter correspond to two rank 2 Fuchsian systems having one regular singular point and one irregular singular point of Poincaré rank $\frac12$ which can be expicitly solved in terms of Bessel functions. 
 
 Let us also remark that (i) only $\hat\Psi^{[0]}\lb z\rb$ depends on PIII ($D_8$) independent variable $t$ (via the asymptotic condition at $z=0$);
 (ii) the initial RHP may also be rewritten as a RHP on a single circle inside $\mathcal A$ with the jump $\hat\Psi^{[0]}\lb z\rb
 {\hat{\Psi}^{[\infty]}\lb z\rb}^{-1}$. The study of an equivalent RHP on a circle is the main tool used in \cite{Niles} for the asymptotic analysis of PIII ($D_8$).
 
 \subsection{Building block solutions}
 
 Consider a model differential system  
 \beq\label{modelSinfty}
 \partial_{z}Y^{[\infty]}= A^{[\infty]}\lb z\rb Y^{[\infty]},
 \qquad A^{[\infty]}\lb z\rb=-\sigma_+ + 
 \lb\begin{array}{cc}
 \sigma+\frac{1}{2} & 0 \\
 -1 & -\sigma-\frac{1}{2} \end{array}\rb z^{-1} .
 \eeq
 Such an ansatz is inspired by the following: we would like to have an irregular singularity of Poincaré rank $\frac12$ at $z=\infty$ and a regular singularity at $z=0$ with local monodromy exponents given by 
 $\pm\lb\sigma+\frac12\rb$, cf (\ref{eigenM0}). 
 
 Choose the fundamental matrix solution of (\ref{modelSinfty}) as
 \beq\label{fundinfty}
 Y^{[\infty]}_{\infty}\lb z\rb=\frac{1}{i\sqrt{2\pi}}\lb\begin{array}{cc}
 2\sqrt z\,K_{2\sigma}\lb2\sqrt z\rb & 
 2\pi\sqrt z\,I_{2\sigma}\lb2\sqrt z\rb\vspace{0.1cm}\\
 2K_{2\sigma+1}\lb2\sqrt z\rb
 & -2\pi I_{2\sigma+1}\lb 2\sqrt z\rb
 \end{array}\rb 
 \lb\begin{array}{cc} 1 & - ie^{ 2\pi i\sigma} \\ 0 & 1\end{array}\rb,
 \eeq
 where $I_{\nu}\lb x\rb $, $K_{\nu}\lb x\rb$ denote the modified Bessel functions of the 1st and 2nd kind. This fundamental solution is defined in the domain $\arg z\in ]0,2\pi[$ where it has the asymptotics 
 \ben
 Y^{[\infty]}_{\infty}\lb z\rb\simeq G\lb\sqrt{z}\rb\left[\mathbb 1+\sum_{k=1}^{\infty}y_k^{[\infty]}z^{-\frac k2} \right] e^{-2\sigma_z \sqrt{z}},\qquad z\to\infty.
 \ebn
 In the vicinity of $z=0$, it becomes convenient to rewrite it as
 \begin{subequations}
 \begin{align}
 \label{Yinfinfa}
 Y^{[\infty]}_{\infty}\lb z\rb=&\,Y^{[\infty]}_{0}\lb z\rb 
 \,e^{2\pi i\eta\sigma_z} z^{\mathfrak S} UE,\\
 Y^{[\infty]}_{0}\lb z\rb=&\,i\sqrt{\frac{\pi}{ \sin2\pi\sigma}}
 \lb\begin{array}{cc}
 z^{-\sigma}I_{2\sigma}\lb2\sqrt z\rb & - z^{\sigma+1}I_{-2\sigma}\lb2\sqrt z\rb\vspace{0.1cm}\\
 -z^{-\sigma-\frac12}I_{2\sigma+1}\lb2\sqrt z\rb & z^{\sigma+\frac12}I_{-2\sigma-1}\lb2\sqrt z\rb\end{array}\rb e^{-i\pi \lb\sigma-\frac14\rb\sigma_z}.
 \end{align}
 \end{subequations}
 In fact, we used  in (\ref{fundinfty}) the same $\sigma$ as in the parameterization of Stokes data precisely to achieve (\ref{Yinfinfa}). The matrix function $Y^{[\infty]}_0\lb z\rb$ is holomorphic and invertible in the entire complex plane, and normalized as to have unit determinant. Therefore, the solution $\hat{\Psi}^{[\infty]}\lb z\rb$ of the exterior auxiliary RHP may be written as
 \beq\label{solpsiext}
 \hat{\Psi}^{[\infty]}\lb z\rb=\begin{cases}
 Y^{[\infty]}_{\infty}\lb z\rb,\qquad & z\text{ outside } C_{\infty},\vspace{0.1cm}\\
 Y^{[\infty]}_{0}\lb z\rb e^{2\pi i\eta\sigma_z},\qquad & z\text{ inside } C_{\infty}.
 \end{cases}
 \eeq
 
 Similarly, the function 
 \beq\label{coef00}
 Y^{[0]}_{0}\lb z\rb=\frac{1}{i\sqrt{2\pi}}\lb\begin{array}{cc}
 2\pi I_{-2\sigma-1}\lb\frac{2}{\sqrt z}\rb &
 2K_{-2\sigma-1}\lb\frac{2}{\sqrt z}\rb \vspace{0.1cm} \\
 \frac{2\pi}{\sqrt z} I_{-2\sigma}\lb\frac{2}{\sqrt z}\rb &
 -\frac{2}{\sqrt z} K_{-2\sigma}\lb\frac{2}{\sqrt z}\rb
 \end{array}\rb \lb\begin{array}{cc}
 1 & 0 \\
 -i e^{2\pi i \sigma}& 1\end{array}\rb
  \eeq
 defines a fundamental matrix solution of the linear system
 \beq\label{modelSzero}
 \partial_z Y^{[0]}=A^{[0]}\lb z\rb Y^{[0]},\qquad
 A^{[0]}\lb z\rb=-\sigma_- z^{-2}+\lb\begin{array}{cc}
 \sigma+\frac12 & -1 \\ 0 & -\sigma-\frac12
 \end{array}\rb z^{-1}.
 \eeq
 It is characterized by the asymptotic behavior 
 \beq\label{formalmodel0}
   Y^{[0]}_{0}\lb z\rb\simeq G\lb\sqrt z\rb\left[\mathbb 1+\sum_{k=1}^{\infty}y^{[0]}_k z^{\frac k2} \right] e^{2\sigma_z/ \sqrt z},
  \eeq
  as $z\to 0$ inside the sector $\arg z\in]0,2\pi[$. In the neighborhood of $z=\infty$, this model solution $Y^{[0]}_0\lb z\rb$ can be suitably rewritten as
  \begin{subequations}
  \begin{align}
  & Y^{[0]}_0\lb z\rb=Y^{[0]}_{\infty}\lb z\rb z^{\mathfrak S}U,\\&
 Y^{[0]}_{\infty}\lb z\rb=i\sqrt{\frac{\pi}{\sin2\pi\sigma}}
 \lb\begin{array}{cc}
 z^{-\sigma-\frac12} I_{-2\sigma-1}\lb\frac{2}{\sqrt z}\rb &
 z^{\sigma+\frac12} I_{2\sigma+1}\lb\frac{2}{\sqrt z}\rb \vspace{0.1cm}\\
 z^{-\sigma-1} I_{-2\sigma}\lb\frac{2}{\sqrt z}\rb &
 z^{\sigma} I_{2\sigma}\lb\frac{2}{\sqrt z}\rb
 \end{array}\rb e^{-i\pi\lb\sigma+\frac14\rb\sigma_z},
  \end{align}
  \end{subequations}
   Taking into account that the matrix ratio ${G\lb \sqrt z\rb}^{-1}G\lb\sqrt{\frac zt}\rb=t^{-\frac{\sigma_x}{4}}$ is independent of $z$, the solution $\hat{\Psi}^{[0]}\lb z\rb$ of the interior auxiliary RHP may now be expressed as
    \beq\label{solpsiint}
    \hat{\Psi}^{[0]}\lb z\rb=\begin{cases}
    Y^{[0]}_{0}\lb \frac{z}{t}\rb,\qquad & z\text{ inside } C_{0},\vspace{0.1cm}\\
    Y^{[0]}_{\infty}\lb \frac zt\rb t^{-\mathfrak S},\qquad &z\text{ outside } C_{0}.
    \end{cases} 
    \eeq

 The parameterization of Stokes data introduced in Subsection~\ref{subsecMonodromy} now becomes more transparent. The   variable $\sigma$ encodes the spectrum of the single nontrivial monodromy matrix $M_0$ whose eigenvalues are given by
     $-e^{\pm 2\pi i \sigma}$, cf (\ref{eigenM0}). The 2nd parameter $\eta$ measures a relative twist of local parametrices $Y^{[\infty]}_0\lb z\rb$, $Y^{[0]}_{\infty}\lb\frac zt\rb$ in the full solution $\hat{\Psi}\lb z\rb$.
 
 \section{Fredholm determinant representation\label{sec_fred}}
 \subsection{Boundary spaces}          
 Let $V(C)$ be the space of smooth functions on a circle $C$ which will be sometimes identified with the space of holomorphic functions in an annulus containing $C$. Also, define the space $H\lb C\rb=\mathbb C^2\otimes V\lb C\rb$ whose elements will be represented as $2$-rows of elements of $V\lb C\rb$. The subspaces of $V\lb C\rb$ and $H\lb C\rb$ that consist of functions with only positive or negative Fourier modes will be denoted by $V_{\pm}\lb C\rb$ and $H_{\pm}\lb C\rb$. 
 
 In relation with the previously discussed RHP for the function $\hat\Psi\lb \xi\rb$, introduce the spaces
 \beq\label{decomp}
 \mathcal H=\mathcal H_+\oplus \mathcal H_-,\qquad \mathcal H_{\pm}=H_{\pm}\lb C_0\rb\oplus H_{\mp}\lb
 C_{\infty}\rb.
 \eeq
 Observe that each of the subspaces $\mathcal H_{\pm}$ can be identified in a natural way with the space of vector-valued holomorphic functions on the annulus $\mathcal A$. We are now going to consider two operators acting on $\mathcal H$ \textit{from the right} and generalizing the usual projections on positive and negative modes.
 \begin{enumerate}
 \item The first operator, to be denoted by $\mathcal P_{\Sigma}$, is defined by 
  \ben
  \lb f\mathcal P_{\Sigma}\rb\lb z\rb = \frac1{2\pi i}\int_{C_0\cup C_{\infty}}\frac{f\lb z'\rb {\hat{\Psi}_+\lb z'\rb}^{-1}\hat\Psi_+\lb z\rb dz'}{z-z'},\qquad z\in 
  C_0\cup C_{\infty}.
  \ebn             
 The convention used to interpret the singularity at $z'=z$ is to slightly deform the integration contour so that it goes clockwise around this point.
 \item The second operator, $\mathcal P_{\oplus}$, is constructed in a similar way with the help of elementary building block solutions $\hat\Psi^{[0]}\lb z\rb$ and $\hat\Psi^{[\infty]}\lb z\rb$,
 \ben
 \lb f\mathcal P_{\oplus}\rb\lb z\rb =
 \frac1{2\pi i}\int_{C_k}\frac{f\lb z'\rb {\hat{\Psi}^{[k]}_+\lb z'\rb}^{-1}\hat\Psi_+^{[k]}\lb z\rb dz'}{z-z'},\qquad\quad z\in 
   C_k,\quad k=0,\infty.
 \ebn 
 \end{enumerate}
 The absence of jumps of $\hat\Psi\lb z\rb{\hat{\Psi}^{[0]}_+\lb z\rb}^{-1}$ (and $\hat\Psi\lb z\rb{\hat{\Psi}^{[\infty]}_+\lb z\rb}^{-1}$) inside  $C_0$ (resp. outside $C_{\infty}$), and systematic application of residue theorem/collapsing the contours imply the following properties:
 \begin{itemize}
 \item $\mathcal P_{\Sigma}^2=\mathcal P_{\Sigma} $ and
 $\mathcal P_{\oplus}^2=\mathcal P_{\oplus} $, i.e.  the operators $\mathcal P_{\Sigma}$, $\mathcal P_{\oplus}$ are projections.
 \item $\operatorname{ker}\mathcal P_{\oplus}=\mathcal H_-$, $\operatorname{ker}\mathcal P_{\Sigma}\supseteq \mathcal H_{\mathcal A}$, where $\mathcal H_{\mathcal A}$ is the space of boundary values of functions holomorphic on $\mathcal A$.
 \item $\mathcal P_{\Sigma}\mathcal P_{\oplus}=\mathcal P_{\Sigma}$,
 $\mathcal P_{\oplus}\mathcal P_{\Sigma}=\mathcal P_{\oplus}$; this means that $\mathcal P_{\Sigma}$ and $\mathcal P_{\oplus}$ have the same range, to be denoted by $\mathcal H_{\mathcal T}$.
 \end{itemize}
 Loosely speaking, the space $\mathcal H_{\mathcal T}$  consists of functions on $C_0\cup C_{\infty}$ whose continuations outside $\mathcal A$  share the global monodromy properties of the fundamental matrix solution of (\ref{LStrans2}). The operators $\mathcal P_{\Sigma}$, $\mathcal P_{\oplus}$ project on $\mathcal H_{\mathcal T}$ along $\mathcal H_{\mathcal A}$ and $\mathcal H_-$, respectively, which may be denoted as $\mathcal P_{\Sigma}=\mathcal H\xrightarrow{\mathcal H_{\mathcal A}}\mathcal H_{\mathcal T}$, $\mathcal P_{\oplus}=\mathcal H\xrightarrow{\mathcal H_{-}}\mathcal H_{\mathcal T}$.
 
 According to the decomposition (\ref{decomp}), write $f\in \mathcal H$  as
 \ben
 f=\left(\begin{array}{cc}
 f^{[0]}_+ & f^{[\infty]}_-\end{array}\right)\oplus
 \left(\begin{array}{cc}
  f^{[0]}_- & f^{[\infty]}_+\end{array}\right).
 \ebn
 The action of $\mathcal P_{\oplus}$ is then given by
 \beq\label{Poplusaction}
 f\mathcal P_{\oplus}=
 \left(\begin{array}{cc}
  f^{[0]}_+ & f^{[\infty]}_-\end{array}\right)\oplus\left(\begin{array}{cc}
   f^{[0]}_+\mathsf{d} & f^{[\infty]}_-\mathsf{a}\end{array}\right),
 \eeq
 where the matrix integral operators $\mathsf{a}: H_-\lb C_{\infty}\rb\to
 H_+\lb C_{\infty}\rb$ and $\mathsf{d}: H_+\lb C_0\rb\to H_-\lb C_0\rb$ are expressed in terms of elementary solutions $\hat\Psi^{[0]}\lb z\rb$, $\hat\Psi^{[\infty]}\lb z\rb$ and have integrable form:
 \beq
 \label{adopsdef}
 \begin{aligned}
 \lb f\mathsf a\rb \lb z\rb=\frac{1}{2\pi i} \oint_{C_{\infty}}
 f\lb z'\rb \mathsf a\lb z',z\rb dz',&\qquad 
  \lb f\mathsf d\rb \lb z\rb=-\frac{1}{2\pi i} \oint_{C_{0}}
  f\lb z'\rb \mathsf d\lb z',z\rb dz',\\
  \mathsf a\lb z',z\rb=\frac{{\hat{\Psi}^{[\infty]}_+\lb z'\rb}^{-1}\hat\Psi_+^{[\infty]}\lb z\rb-\mathbb 1}{z-z'},&\qquad \mathsf d\lb z',z\rb=
  \frac{\mathbb 1-{\hat{\Psi}^{[0]}_+\lb z'\rb}^{-1}\hat\Psi_+^{[0]}\lb z\rb}{z-z'}.
 \end{aligned} 
 \eeq
 The minus sign is introduced into the definition of $\mathsf d\lb z',z\rb$ to absorb the opposite orientation of the circles $C_{0,\infty}$ in some computations below.
 Let us note in passing that the action of $\mathsf a$ and $\mathsf d$ may be extended from the boundary circles to vector-valued functions holomorphic on $\mathcal A$. 
 
  The result (\ref{Poplusaction}) suggests that $f^{[0]}_+$,  $f^{[\infty]}_-$ provide convenient coordinates on $\mathcal H_{\mathcal T}$.  We are going to use this basis to describe the operator $\mathcal P_{\Sigma}$ involving the solution $\hat{\Psi}\lb \xi\rb$. Given $f\in\mathcal H$, write $f=g+h$ with $g\in\mathcal H_{\mathcal T}$ and $h\in\mathcal H_{A}$. These conditions translate into
  \ben
  g=\left(\begin{array}{cc}
    g^{[0]}_+ & g^{[\infty]}_-\end{array}\right)\oplus\left(\begin{array}{cc}
     g^{[0]}_+\mathsf{d} & g^{[\infty]}_-\mathsf{a}\end{array}\right),\qquad
     h=\left(\begin{array}{cc}
       h^{[0]}_+ & h^{[\infty]}_-\end{array}\right)\oplus\left(\begin{array}{cc}
        h^{[\infty]}_- & h^{[0]}_+\end{array}\right)
  \ebn
  Expressing $h^{[0]}_+$, $h^{[\infty]}_-$ in terms of $g^{[0]}_+$, $g^{[\infty]}_-$, one obtains the equation
  \ben
  \left(\begin{array}{cc}
      g^{[0]}_+ & g^{[\infty]}_-\end{array}\right)
      \lb \mathbb 1-K\rb=\left(\begin{array}{cc}
            f^{[0]}_+-f^{[\infty]}_+ & f^{[\infty]}_-
            -f^{[0]}_-\end{array}\right),
            \qquad
      K=\left(\begin{array}{cc}
      0 & \mathsf a \\ \mathsf d & 0\end{array}\right).
  \ebn
  Below we assume invertibility of $\mathbb 1-K$, which ensures the existence of a unique splitting $\mathcal H=\mathcal H_{\mathcal T}\oplus \mathcal H_{\mathcal A}$.
  Computing the action of $\mathcal P_{\Sigma}$ on $\mathcal H$ (essentially equivalent to solving the original RHP) thereby  amounts to finding the inverse $\lb \mathbb 1-K\rb^{-1}$.
  
  Consider the restrictions $\mathcal P_{\oplus,+}=\mathcal P_{\oplus}\bigl|_{\mathcal H_+}$,  $\mathcal P_{\Sigma,+}=\mathcal P_{\Sigma}\bigl|_{\mathcal H_+}$. We have just seen that in the previously described basis $\mathcal P_{\oplus,+}$ is given by the identity matrix, whereas
  $\mathcal P_{\Sigma,+}$ coincides with $\lb \mathbb 1-K\rb^{-1}$.
  \begin{defin}\label{deftu} The tau function of the Riemann-Hilbert problem for $\hat{\Psi}\lb\xi\rb$ is defined as
  \beq\label{tauRHPdef}
  \tau\lb t\rb=\operatorname{det}\lb\mathcal H_+\xrightarrow{\mathcal H_{-}}
  \mathcal H_{\mathcal T}\xrightarrow{\mathcal H_{\mathcal A}}\mathcal H_{\mathcal +}\rb=\operatorname{det}\lb \mathcal P_{\oplus,+}
  {\mathcal P_{\Sigma,+}}^{-1}\rb = \operatorname{det}\lb\mathbb 1-K\rb.
  \eeq
  \end{defin}
  The first two expressions of $\tau\lb t\rb$ are ``coordinate-free'' while the last Fredholm determinant corresponds to the choice of a specific basis. Our next task is to understand the relation between the last definition and the tau function of Painlevé III ($D_8$) equation introduced in (\ref{deftau}).

  \subsection{Relation to $\tau_{\text{III}}\lb t\rb$}
  Let $t_0$ be a constant parameter close to Painlevé III ($D_8$) independent variable $t$ and consider the ratio
  \begin{align*}
  \frac{\tau\lb t\rb}{\tau\lb t_0\rb}=&\,\operatorname{det}\lb\mathcal H_+\xrightarrow{\mathcal H_{-}}
    \mathcal H_{\mathcal T}\lb t\rb\xrightarrow{\mathcal H_{\mathcal A}}\mathcal H_{\mathcal +}
    \xrightarrow{\mathcal H_{\mathcal A}}\mathcal H_{\mathcal T}\lb t_0\rb\xrightarrow{\mathcal H_{-}}\mathcal H_{\mathcal +}\rb=\\
    =&\,
    \operatorname{det}\lb \mathcal H_{\mathcal T}\lb t_0\rb\xrightarrow{\mathcal H_{-}}\mathcal H_{\mathcal T}\lb t\rb
 \xrightarrow{\mathcal H_{\mathcal A}}\mathcal H_{\mathcal T}\lb t_0\rb   \rb=\\
 =&\,
 \operatorname{det}\lb \mathcal P_{\oplus}\lb t\rb\bigl|_{\mathcal H_{\mathcal T}\lb t_0\rb}\mathcal P_{\Sigma}\lb t_0\rb\bigl|_{\mathcal H_{\mathcal T}\lb t\rb}\rb.
  \end{align*}
  Since for $\nu=\oplus,\Sigma$ we can express the inverses as $\lb\mathcal P_{\nu}\lb t\rb\bigl|_{\mathcal H_{\mathcal T}\lb t_0\rb}\rb^{-1}=\mathcal P_{\nu}\lb t_0\rb\bigl|_{\mathcal H_{\mathcal T}\lb t\rb}$, the logarithmic derivative of $\tau\lb t\rb$ may be written as
  \begin{align}
  \nonumber\partial_t\ln\tau\lb t\rb=&\,-\operatorname{Tr}_{\mathcal H_{\mathcal T}\lb t_0\rb}\left\{
  \mathcal P_{\oplus}\lb t\rb\bigl|_{\mathcal H_{\mathcal T}\lb t_0\rb}\mathcal P_{\Sigma}\lb t_0\rb\bigl|_{\mathcal H_{\mathcal T}\lb t\rb} \partial_t\lb \mathcal P_{\Sigma}\lb t\rb\bigl|_{\mathcal H_{\mathcal T}\lb t_0\rb}\mathcal P_{\oplus}\lb t_0\rb\bigl|_{\mathcal H_{\mathcal T}\lb t\rb}\rb
  \right\}=\\
  \nonumber=&\,-\operatorname{Tr}_{\mathcal H}\Bigl\{
   \mathcal P_{\oplus}\lb t\rb\mathcal P_{\Sigma}\lb t_0\rb
  \partial_t \mathcal P_{\Sigma}\lb t\rb \mathcal P_{\oplus}\lb t_0\rb
  \Bigr\}=\\
  \label{taustep1}=&\,-\operatorname{Tr}_{\mathcal H}\Bigl\{
     \mathcal P_{\oplus}\lb t\rb
    \partial_t \mathcal P_{\Sigma}\lb t\rb 
    \Bigr\}.
  \end{align}
  Here the middle line is obtained by using that $\operatorname{ran}\,\mathcal P_{\nu}\lb t\rb=\mathcal H_{\mathcal T}\lb t\rb$. The last line follows from the transversality of $\mathcal H_{\mathcal T}\lb t\rb$ and $\mathcal H_{\mathcal A}$ (as well as $\mathcal H_{\mathcal T}\lb t\rb$ and $\mathcal H_{-}$) in $\mathcal H$, which implies that
  \ben
  \lb\mathcal H\xrightarrow{\mathcal H_{\mathcal T}\lb t_0\rb} \mathcal H_{\mathcal A}\xrightarrow{\mathcal H_{\mathcal A}} \mathcal H_{\mathcal T}\lb t\rb\rb\quad=
  \quad
  \lb\mathcal H\xrightarrow{\mathcal H_{\mathcal T}\lb t_0\rb} \mathcal H_{-}\xrightarrow{\mathcal H_{-}} H_{\mathcal T}\lb t\rb\rb
  \quad =\quad0,
  \ebn
  i.e. the corresponding compositions of projections are equal to zero.
  
  The next task is to compute the trace in the right side of (\ref{taustep1}). Collapsing the contours and computing residues as in Step 2 of the proof of Theorem~2.9 in \cite{GL16}, we arrive at
  \ben
  \operatorname{Tr}_{\mathcal H}\Bigl\{
       \mathcal P_{\oplus}\lb t\rb
      \partial_t \mathcal P_{\Sigma}\lb t\rb 
      \Bigr\}=\sum_{\nu=0,\infty}\frac{1}{2\pi i}
      \oint_{C_{\nu}} \operatorname{Tr}\left\{
      \partial_{z}\lb\hat\Psi_+\lb z\rb
      {\hat\Psi^{[\nu]}_+\lb z\rb}^{-1}\rb \hat\Psi^{[\nu]}_+\lb z\rb\partial_t\lb {\hat\Psi_+\lb z\rb}^{-1}\rb\right\}dz.
  \ebn
  Recall that $\hat\Psi$ has the same jumps as $\hat\Psi^{[0]}$ inside $C_0$ and as $\hat\Psi^{[\infty]}$ outside $C_{\infty}$. Therefore the ``$+$''-indices in the above expression are redundant, the contours $C_{0,\infty}$ can be replaced by small circles around $0$ and $\infty$, and the resulting integrals may be computed by residues. On these circles, the integrand may be represented   by series involving only integer (but not half-integer) powers of $z$, which can be shown using once again the symmetry properties such as $G\lb \xi e^{i\pi}\rb=iG\lb\xi\rb\sigma_x$ and  $\sigma_x \tilde Y_{\mathrm{form}}^{(0)}\lb -\xi\rb\sigma_x=\tilde Y_{\mathrm{form}}^{(0)}\lb\xi\rb$. Furthermore,
 the series at~$\infty$ has the form $\sum_{k\ge 0} f_k z^{-k-2}$,  hence the corresponding residue  vanishes. 
 
 On the other hand, the residue at $0$ reads (note the negative orientation of $C_0$)
 \ben
 \frac{\lb y^{(0)}_1\rb_{11}-\lb y^{(0)}_1\rb_{12}}{\sqrt t}-
 \frac{\lb y^{[0]}_1\rb_{11}-\lb y^{[0]}_1\rb_{12}}{ t},
 \ebn
 where $y^{(0)}_1$ is the first nontrivial coefficient of the formal solution (\ref{formalsol0}) and $y^{[0]}_1$ is its counterpart in the expansion (\ref{formalmodel0}) of the model solution $Y^{[0]}_0\lb z\rb$. The former quantity is explicitly given by (\ref{expcoef1}), while the latter is readily deduced from
 (\ref{coef00}):
 \ben
 y^{[0]}_1=\lb\sigma+\frac14\rb\frac{i\sigma_y}{2}-\lb \sigma+\frac14\rb^2\sigma_z.
 \ebn
 Combining the two results with (\ref{taustep1}) yields
 \ben
 \partial_t\ln\tau\lb t\rb=-\operatorname{Tr}_{\mathcal H}\Bigl\{
        \mathcal P_{\oplus}\lb t\rb
       \partial_t \mathcal P_{\Sigma}\lb t\rb 
       \Bigr\}=\frac1t\left[\frac{p^2}{q^2}-q-\frac tq-\lb\sigma+\frac12\rb^2\right].
 \ebn
 
 The Fredholm determinant $\tau\lb t\rb$ from the Definition~\ref{deftu} may therefore be identified with the usual  Pain\-levé~III ($D_8$) tau function $\tau_{\mathrm{III}}\lb t\rb$ defined by (\ref{deftau}):
  \beq
  \tau_{\mathrm{III}}\lb t\rb= \mathrm{const}\cdot t^{
  \lb\sigma+\frac12\rb^2}\tau\lb t\rb.
  \eeq
 In combination with explicit solutions (\ref{solpsiext}), (\ref{solpsiint}) of auxiliary RHPs which appear in the definition (\ref{adopsdef}) of operators $\mathsf a$ and $\mathsf d$, this yields the following result.
 \begin{theo}\label{theoFr} Let $\lb \sigma,\eta\rb\in \mathbb C^2$ with $\sigma\notin\mathbb Z/2$ be the coordinates on the generic stratum of the space of the Stokes data of the linear system (\ref{LSsub}), introduced  in Subsection~\ref{subsecMonodromy}. The corresponding Pain\-levé~III ($D_8$) tau function $\tau_{\mathrm{III}}\lb t\rb=\tau_{\mathrm{III}}\lb t\,|\,\sigma,\eta\rb$
 can be expressed as Fredholm determinant
  \beq\label{blockFr}
  \tau_{\mathrm{III}}\lb t\rb= \mathrm{const}\cdot t^{
  \lb\sigma+\frac12\rb^2}\operatorname{det}\lb \mathbb 1-K\rb,\qquad
        K=\left(\begin{array}{cc}
        0 & \mathsf a \\ \mathsf d & 0\end{array}\right).
  \eeq
  Here the operators $\mathsf a$, $\mathsf d$ act on vector-valued functions $f\in H\lb C\rb$ on a circle $C$ centered at the origin and oriented counterclockwise,
  \beq\label{adopsdef2}
   \lb f\mathsf a\rb \lb z\rb=\frac{1}{2\pi i} \oint_{C}
   f\lb z'\rb \mathsf a\lb z',z\rb dz',\qquad 
    \lb f\mathsf d\rb \lb z\rb=\frac{1}{2\pi i} \oint_{C}
    f\lb z'\rb \mathsf d\lb z',z\rb dz',
  \eeq
  and the integral kernels
  $\mathsf a\lb z',z\rb$, $\mathsf d\lb z',z\rb$ are explicitly given by
   \begin{subequations}\label{adbessel}
  \begin{align}
  \label{abessel}
  & \mathsf a\lb z',z\rb  =\;e^{i\pi \lb \sigma-2\eta\rb\sigma_z}\;\frac{\mathbb J_{\sigma}\lb z',z\rb-\mathbb 1}{z-z'}\; e^{i\pi  \lb 2\eta-\sigma\rb\sigma_z} ,\\
 \label{38b} & \mathsf d\lb z',z\rb   =t^{\mathfrak S}e^{i\pi\sigma\sigma_z}\sigma_y \frac{\mathbb 1-\mathbb J_{\sigma}\lb \frac t{z'},\frac t{z}\rb}{z-z'}\,\sigma_y e^{-i\pi\sigma\sigma_z}
  t^{-\mathfrak S},\\
  &\mathbb{J}_{\sigma}\lb z',z\rb= \frac{\pi}{\sin 2\pi\sigma}
    \lb\begin{array}{cc}
    z'j_{\sigma+\frac12}(z)
      j_{-\sigma}(z')-j_{\sigma}(z) j_{-\sigma-\frac12}(z') & 
   iz'j_{-\sigma-\frac12}(z)j_{-\sigma}(z')-izj_{-\sigma}(z)j_{-\sigma-\frac12}(z')\vspace{0.1cm}\\
   i j_{\sigma+\frac12}(z) j_{\sigma}(z')- ij_{\sigma}(z)j_{\sigma+\frac12}(z') &
   zj_{-\sigma}(z)j_{\sigma+\frac12}(z')-
   j_{-\sigma-\frac12}(z)j_{\sigma}(z')
    \end{array}\rb,
  \end{align}
  \end{subequations}
  with
  $j_{\sigma}(z)=z^{-\sigma}I_{2\sigma}\lb 2\sqrt z\rb=\ds\frac{_0F_1\lb 2\sigma+1;z\rb}{\Gamma\lb 2\sigma+1\rb}$ and $\mathfrak S=\lb\sigma+\frac12\rb\sigma_z$.
 \end{theo}
   
  \begin{rmk} The kernels $\mathsf a\lb z',z\rb$, $\mathsf d\lb z',z\rb$ are not singular at $z=z'$.
  That $j_{\sigma}(z)$ are holomorphic in the entire complex plane is a signature of the fact that $\operatorname{ran}\mathsf a\subseteq H_+\lb C\rb\subseteq\operatorname{ker}\mathsf a$ and $\operatorname{ran}\mathsf d\subseteq H_-\lb C\rb\subseteq\operatorname{ker}\mathsf d$. The Fredholm determinant may therefore be rewritten as 
  \ben
  \operatorname{det}\lb \mathbb 1-K\rb=
  \operatorname{det}\lb \mathbb 1+\mathsf a+\mathsf d\rb. 
  \ebn
  The latter form may seem more compact while the  integral kernel of $\mathsf a+\mathsf d$ is still integrable. However it turns out to be beneficial for our purposes to work with the block structure of $K$ in (\ref{blockFr}).
  \end{rmk} 
    \begin{rmk}
    Let us note that the tau function (\ref{deftau}) differs from \cite[Eq. (2.14)]{GIL13} or \cite[Eq.~(2.15)]{ILT14} by a $\mathbb Z_2$-B\"acklund transformation. This discrete symmetry becomes most explicit at the level of the sine-Gordon equation (\ref{rSG}) where it corresponds to the mapping $u\mapsto -u$. The relevant monodromy parameters transform as $\sigma\mapsto \frac12-\sigma$, $\eta\mapsto -\eta $ which should be taken into account before comparing (\ref{blockFr})--(\ref{adbessel}) with
    (\ref{piiicft}). An interesting representation-theoretic interpretation of this symmetry has been recently suggested in \cite{BSh2}.
    \end{rmk}
  \begin{rmk}\label{asymrmk}
  The monodromy data $\bigl\{S^{(\nu)}_k\bigr\}$, $E$ in (\ref{monodrparam}) are invariant with respect to integer shifts $\sigma\mapsto \sigma+1$. The Painlevé III ($D_8$) tau function should thus be quasi-periodic in $\sigma$, namely, 
  \ben
  \tau_{\mathrm{III}}\lb t\,|\,\sigma+1,\eta\rb=\operatorname{const}\cdot \tau_{\mathrm{III}}\lb t\,|\,\sigma,\eta\rb , 
  \ebn
  where the constant expression depends on the choice of normalization of $\tau_{\mathrm{III}}\lb t\,|\,\sigma,\eta\rb$. This quasi-periodicity is not obvious at all at the level of Fredholm determinant representation (\ref{blockFr})--(\ref{adbessel}) but will be made manifest in the next section. 
  
  Upon truncation of the Taylor expansion  of the right side of (\ref{38b}) in $t$, the operator $\mathsf d$ becomes finite rank so that the corresponding $t\to 0$ asymptotics of $\tau_{\mathrm{III}}\lb t\rb$  is given by a finite determinant, cf \cite[Theorem 2.11]{GL16}. 
  From the point of view of this asymptotic analysis, the most  efficient choice of $\sigma$ is to set $-1< \Re\sigma\leq0$. The leading asymptotic terms obtained by such procedure coincide with the known results \cite{Jimbo82,IN,Novokshenov,FIKN,Niles}. It is an instructive exercise to check that the subleading asymptotic terms derived from the Fredholm determinant reproduce \cite[Eqs. (3.3)--(3.5)]{ILT14}.
  \end{rmk}
  \begin{rmk}
  The Painlevé III ($D_8$) isomonodromic RHP is usually formulated in the literature for the unfolded system
  (\ref{LStrans2}), see e.g. \cite{FIKN,IP,Niles}. While such formulation has a number of technical advantages, the correspondence with CMR confluence diagram is not manifest therein. Furthermore, the analog of the Fredholm determinant (\ref{tauRHPdef}) does \textit{not} coincide with the tau function (\ref{deftau}). Instead, it gives the tau function of a special case of PIII~($D_6$) equation ($N_f=2$ in the gauge theory language), related to
  PIII ($D_8$) by a quadratic transformation.
  \end{rmk}

 \section{Series over Young diagrams}
 \subsection{Cauchy matrix representations}
 Let us now express the operators $\mathsf a$ and $\mathsf d$ from (\ref{adopsdef2}) in the basis of Fourier modes, where they are given by semi-infinite matrices. Denoting  $\mathbb Z'=\mathbb Z -\frac12$,
 $\mathbb Z'_+=\mathbb N-\frac12$, write
 \begin{subequations}
 \begin{align}
 \label{fourieraop}
 \mathsf a\lb z',z\rb =&\,\sum_{p,q\in\mathbb Z'_+}\mathsf{a}^{\;\;\; p}_{-q}z'^{-\frac12+p}z^{-\frac12+q},\\
 \mathsf d\lb z',z\rb =&\,\sum_{p,q\in\mathbb Z'_+}\mathsf{d}^{- p}_{\;\;\;q}z'^{-\frac12-p}z^{-\frac12-q},
 \end{align}
 \end{subequations}
 where $z',z\in\mathbb C^*$. The mode operators  $\mathsf{a}^{\;\;\; p}_{-q},\mathsf{d}^{- p}_{\;\;\;q}$ are $2\times 2$ matrices whose elements will be represented as $\mathsf{a}^{\;\;\;\, p;s'}_{-q;s}$, $\mathsf{d}^{- p;s'}_{\;\;\;\,q;s}$, with ``color'' indices $s',s\in\{+,-\}$. Our convention is that ``$+$'' and ``$-$'' correspond to the first and second row/column.
 
 In order to compute these matrix elements explicitly, let us return to the original definition (\ref{adopsdef}) of $\mathsf a$ and $\mathsf d$. Recall that inside the annulus $\mathcal A$ we have
 $\hat{\Psi}^{[\infty]}\lb z\rb=Y^{[\infty]}_{\infty}\lb z\rb 
 E^{-1}U^{-1}z^{-\mathfrak S}$ and $\hat{\Psi}^{[0]}\lb z\rb= Y^{[0]}_{0}\lb\frac zt\rb U^{-1}z^{-\mathfrak S}$, where $Y^{[\infty]}_{\infty}\lb z\rb $, $Y^{[0]}_{0}\lb z\rb$ solve the linear systems (\ref{modelSinfty}), (\ref{modelSzero}). These relations may be used to differentiate the kernels $\mathsf a\lb z',z\rb$, $\mathsf d\lb z',z\rb$ with respect to their arguments. In particular, for $z',z\in\mathcal A$ one has
 \begin{align*}
 &\lb z\partial_z+z'\partial_{z'}+1\rb \mathsf a\lb z',z\rb-
 \left[\mathfrak S,\mathsf a\lb z',z\rb\right]=\\ &\qquad\qquad
= {\hat{\Psi}^{[\infty]}\lb z'\rb}^{-1}
 \frac{zA^{[\infty]}\lb z\rb-z'A^{[\infty]}\lb z'\rb}{z-z'}\hat{\Psi}^{[\infty]}\lb z\rb
  =- {\hat{\Psi}^{[\infty]}\lb z'\rb}^{-1}\sigma_+\hat{\Psi}^{[\infty]}\lb z\rb
  = \\ &\qquad\qquad=-e^{-2\pi i \eta\sigma_z}{Y^{[\infty]}_0\lb z'\rb}^{-1}
 \lb\begin{array}{c} 1 \\ 0\end{array}\rb\otimes 
 \lb\begin{array}{cc} 0 & 1\end{array}\rb
 Y^{[\infty]}_0\lb z\rb e^{2\pi i \eta\sigma_z}=\\
 &\qquad\qquad =-\frac{\pi}{\sin2\pi\sigma}
 \lb\begin{array}{c} f_{-\sigma-\frac12}\lb z'\rb\vspace{0.1cm} \\
 i e^{2\pi i \lb 2\eta-\sigma\rb}f_{\sigma+\frac12}\lb z'\rb 
 \end{array}\rb \otimes
 \lb\begin{array}{cc}
 f_{\sigma+\frac12}\lb z\rb & 
 ie^{2\pi i \lb \sigma-2\eta\rb}f_{-\sigma-\frac12}\lb z\rb
 \end{array}\rb.
 \end{align*}
 Substituting into the last equation the Fourier representation (\ref{fourieraop}) and using the factorization of the right hand side, we obtain
 \beq\label{eqforapq}
 \lb p+q\rb \mathsf{a}^{\;\;\; p}_{-q}-\left[\mathfrak S,\mathsf{a}^{\;\;\; p}_{-q}\right]=e^{i\pi\lb\sigma- 2\eta\rb\sigma_z}\psi^{p}\lb \nu\rb\otimes 
 \bar\psi_{q}\lb\nu\rb e^{i\pi\lb 2\eta-\sigma\rb\sigma_z},
 \eeq
 with
 \begin{subequations}
 \begin{align}
 \psi^{p;s}\lb\nu\rb=&\,
 \sqrt{\frac{\Gamma\lb1+2s\nu\rb}{
  \Gamma\lb1-2s\nu\rb}}\frac{e^{-i\pi s/4}}{\lb p-\frac12\rb!
 \lb1-2s\nu\rb_{p-\frac12}},\\
  \bar\psi_{p;s}\lb\nu\rb=&\,\sqrt{\frac{\Gamma\lb1-2s\nu\rb}{
    \Gamma\lb1+2s\nu\rb}}\quad
    \frac{e^{i\pi s/4}}{\lb p-\frac12\rb!
  \lb 2s\nu\rb_{p+\frac12}},
 \end{align}
 \end{subequations}
 where $s=\pm$, $\lb\alpha\rb_k=\alpha\lb\alpha+1\rb\ldots \lb\alpha+k-1\rb$ denotes the Pochhammer symbol, and we have introduced instead of $\sigma$ a shifted monodromy parameter $\nu=\sigma+\frac12$ to make the resulting expressions more symmetric. Further introducing shifted momenta
 \ben
\qquad x_{p;s}=p-s\nu,\qquad\qquad p\in\mathbb Z',\; s=\pm,
 \ebn
 the solution of (\ref{eqforapq}) can be written as
 \begin{subequations}\label{cauchyad}
 \beq\label{apq}
 \mathsf{a}^{\;\;\;\, p;s'}_{-q;s}=\frac{\psi^{p;s'}\lb\nu\rb
 \bar\psi_{q,s}\lb\nu\rb}{x_{p;s'}-x_{-q;s}}
 \, e^{i\pi\lb 2\eta-\sigma\rb\lb s-s'\rb},
 \eeq
 where $p,q\in\mathbb Z'_+$, $s',s=\pm1$. We thus conclude that in the Fourier basis the operator $\mathsf a$ is given, up to left and right diagonal factors, by a Cauchy matrix $M_{jk}=\frac{1}{x_j-y_k}$. This allows, {\it inter alia}, to compute any minor of $\mathsf a$ in a factorized form.
  
  The matrix elements of $\mathsf d$ may be computed in a similar fashion, or alternatively deduced by comparison of (\ref{abessel}) and (\ref{38b}). The result is again a Cauchy matrix,
  \beq\label{dpq}
  \mathsf{d}^{- q;s}_{\;\;\;\,p;s'}=\frac{\psi^{q;s}\lb-\nu\rb
   \bar\psi_{p,s'}\lb-\nu\rb}{x_{p;s'}-x_{-q;s}}
   \,  e^{i\pi\sigma\lb s-s'\rb}t^{\lb s-s'\rb\nu + p+q},   
  \eeq
  \end{subequations}
  and its nontrivial part coincides with that of (\ref{dpq}) after replacement $\nu\to -\nu$. The dependence on PIII variable $t$  is isolated in the diagonal factors; cf Remark~\ref{asymrmk}.
  
  \subsection{Maya and Young diagrams\label{subsecmaya}}
  Given a matrix $A\in \operatorname{Mat}_{m\times m}\lb \mathbb C\rb$,
  the von Koch's formula
  \ben
  \mathrm{det}\lb\mathbb 1+ A\rb=\sum_{n=0}^{\infty}\;\sum_{i_1<\ldots<i_n}\operatorname{det} \lb A_{i_ji_k}\rb_{j,k=1}^{\; n}
  \ebn
  expresses the determinant $\mathrm{det}\lb\mathbb 1+ A\rb$ as the sum of principal minors of $A$. While this series of course terminates at $n=m$, the formula has a straightforward generalization to infinite matrices. If $A$ is indexed by elements of a discrete set $\mathfrak X$ instead of $\{1,\ldots,m\}$, then 
  \beq\label{vonK}
  \mathrm{det}\lb\mathbb 1+ A\rb=\sum_{\mathfrak Y\in 2^{\mathfrak X}}\operatorname{det}A_{\mathfrak Y}, 
  \eeq
  where the sum is taken over all subsets $\mathfrak Y$ of $\mathfrak X$ and $A_{\mathfrak Y}$ is the principal minor of $A$ obtained by choosing the rows and columns labeled by $\mathfrak Y$.

              \begin{figure}[h!]
                  \centering
                  \includegraphics[height=5cm]{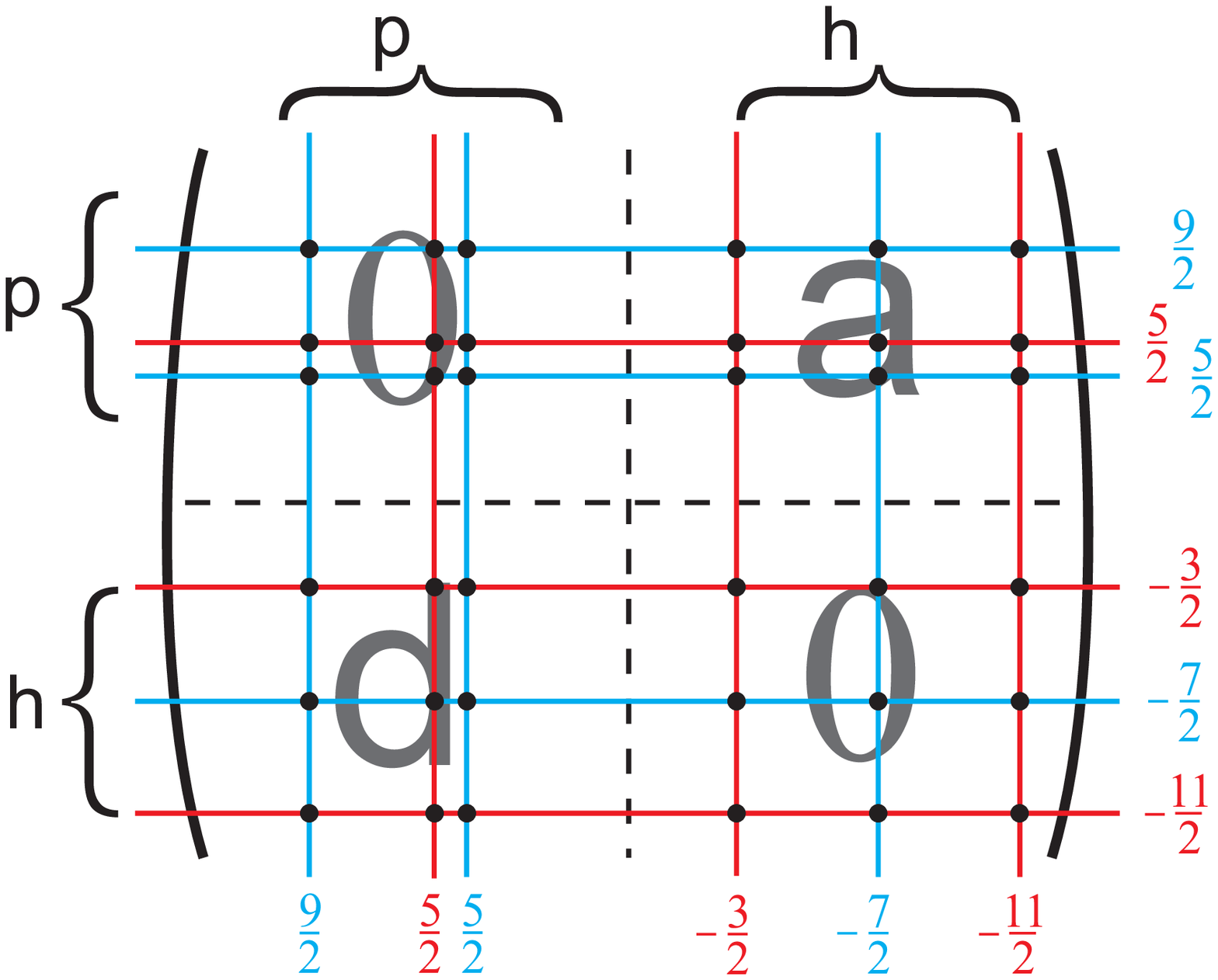}
                  \begin{minipage}{0.75\textwidth}\caption{\label{minselect} Labeling of principal minors of $K$ by positions $\lb\mathsf p,\mathsf h\rb$ of particles and holes of color $+$ (red) and $-$ (blue). Here $\mathsf p^+=\left\{\frac52\right\}$, $\mathsf h^+=\left\{-\frac32,-\frac{11}{2}\right\}$, $\mathsf p^-=\left\{\frac92,\frac52\right\}$, $\mathsf h^-=\left\{-\frac72\right\}$, so that $\mathsf m^+=\left\{\frac52,-\frac32,-\frac{11}{2}\right\}$, $\mathsf m^-=\left\{\frac92,\frac52,-\frac{7}{2}\right\}$
                  and $Q\lb\mathsf m^+\rb=-Q\lb\mathsf m^-\rb=-1$.}
                  \end{minipage}
                  \end{figure}
  
\begin{figure}[h!]
                 \centering
                   \includegraphics[height=4cm]{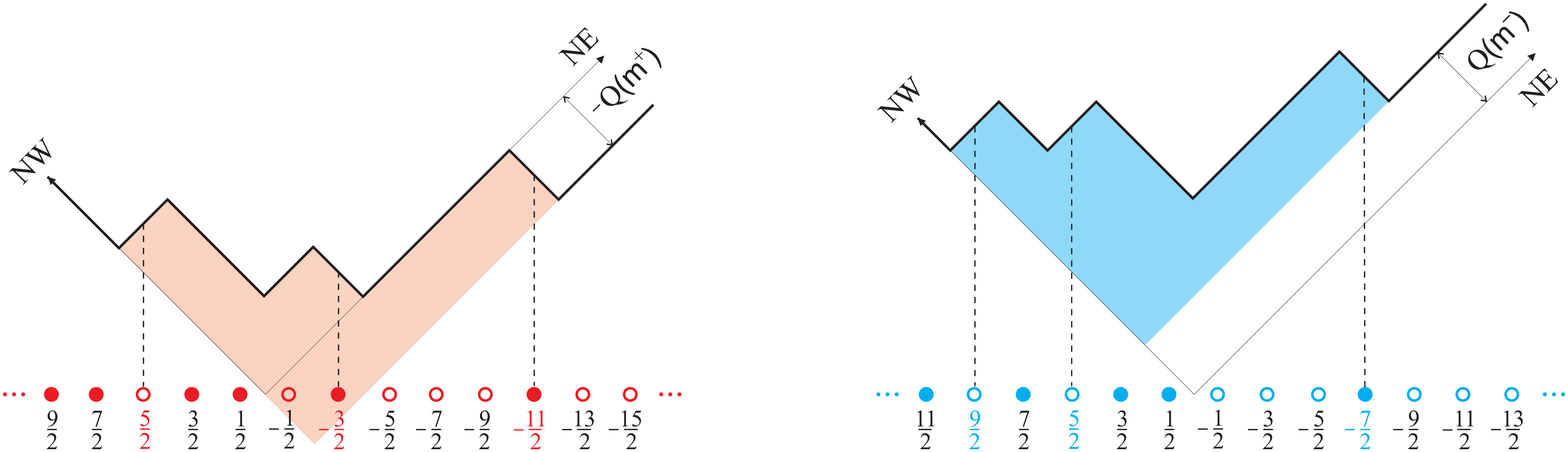}
                   \begin{minipage}{0.65\textwidth}\caption{\label{MayaYoungPIII} Young diagrams $\mathsf Y^+$ and $\mathsf Y^-$ (shaded regions) associated to  $\mathsf m^+=\left\{\frac52,-\frac32,-\frac{11}{2}\right\}$  and $\mathsf m^-=\left\{\frac92,\frac52,-\frac{7}{2}\right\}$.  }
                   \end{minipage}
                   \end{figure}

  We are going apply the last formula to the Fredholm determinant
  (\ref{blockFr}) with $K$ written in the Fourier basis. Represent appropriate subsets as $\mathfrak Y =\lb \mathsf p, \mathsf h\rb$, where $\mathsf p$ and $\mathsf h$ correspond,  respectively, to the first and second block of $K$, see Fig.~\ref{minselect}. The sum in (\ref{vonK}) may be restricted to $(\mathsf p, \mathsf h)$ with $\sharp\lb\mathsf p\rb=\sharp\lb \mathsf h\rb$, as otherwise the corresponding minors obviously vanish. It follows that
  \beq\label{vonKK}
  \operatorname{det}\lb \mathbb 1-K\rb= \sum_{\lb \mathsf p, \mathsf h\rb \,:\, \sharp\lb \mathsf p\rb=\sharp\lb \mathsf h\rb}\lb-1\rb^{\sharp\lb\mathsf p\rb}\operatorname{det}\mathsf a^{\,\mathsf p}_{\,\mathsf h}\operatorname{det}\mathsf d^{\;\mathsf h}_{\;
  \mathsf p},
  \eeq
  where e.g. $\mathsf a^{\,\mathsf p}_{\,\mathsf h}$ is a square $\sharp\lb\mathsf p\rb\times \sharp\lb \mathsf p\rb$ matrix obtained by restricting $\mathsf a$ to rows $\mathsf p$ and columns $\mathsf h$.  
    Let us now take a closer look at the structure of subsets $(\mathsf p, \mathsf h)$ labeling different contributions to (\ref{vonKK}):
  \begin{itemize}  
  \item   The set $\mathsf p$ has the form $\mathsf p^+\sqcup\, \mathsf p^-$, where $\mathsf p^+=\left\{p^+_1,\ldots,p^+_L\right\}$, $\mathsf p^-=\left\{p^-_1,\ldots,p^-_M\right\}$, and $p^{\pm}_j\in\mathbb Z'_+$ are Fourier indices of elements of $\mathsf p$ of color $\pm$. Similarly, $\mathsf h=\mathsf h^+\sqcup\, \mathsf h^-$, where $\mathsf h^+=\left\{-q^+_1,\ldots,-q^+_{L'}\right\}$, $\mathsf h^-=\left\{-q^-_1,\ldots,-q^-_{M'}\right\}$ with $q^{\pm}_j\in\mathbb Z'_+$ consist of Fourier indices of elements of $\mathsf h$ of colors $+$ and $-$. The elements of $\mathsf p^{\pm}$ and $\mathsf h^{\pm}$ are thus distinct positive (resp. negative) half-integers.
  \item Let us consider the combinations $\mathsf m^{\pm}=\mathsf p^{\pm}\cup \mathsf h^{\pm}$. Both $\mathsf m^{\pm}$ are finite subsets of $\mathbb Z'$ and can be represented in the usual way by Maya diagrams; $\mathsf p^{\pm}$ and $\mathsf h^{\pm}$ are positions of particles and holes of color $\pm$ in the Dirac sea, see Fig.~\ref{minselect} and bottom part of Fig.~\ref{MayaYoungPIII}. Given a Maya diagram $\mathsf m$, the difference $Q\lb\mathsf m\rb= \sharp\lb \text{particles}\rb-  \sharp\lb \text{holes}\rb$ is called the charge of $\mathsf m$. The constraint $\sharp\lb \mathsf p\rb=\sharp\lb \mathsf h\rb$ is then nothing but the neutrality condition $Q\lb \mathsf m^+\rb+Q\lb \mathsf m^-\rb=0$. 
  \item On the other hand, the set $\mathbb M$ of Maya diagrams can be bijectively mapped to the set $\mathbb Y\times\mathbb Z$ of charged Young diagrams/partitions. This correspondence is represented graphically in Fig.~\ref{MayaYoungPIII}. The profile of the Young diagram $\mathsf Y\in\mathbb Y$ associated to a Maya diagram $\mathsf m\in\mathbb M$ is obtained by starting far away on the NW-axis and  going south-east above each filled circle and north-east above each empty circle of $\mathsf m$. The charge corresponds to relative position of the bottom boundary of $\mathsf Y$ and the NE-axis, and coincides with $Q\lb\mathsf m\rb$.
  \end{itemize}
 Different contributions to (\ref{vonKK}) may therefore be labeled
 (i) by positions of particles $\mathsf p^{\pm}\in 2^{\mathbb Z'_+}$ and holes $\mathsf h^{\pm}\in 2^{-\mathbb Z'_+}$ of two colors $\pm$ satisfying the balance condition $\sharp\lb\mathsf p^+\rb+\sharp\lb\mathsf p^-\rb=\sharp\lb\mathsf h^+\rb+\sharp\lb\mathsf h^-\rb$; (ii) by pairs $\lb\mathsf m^+,\mathsf m^-\rb\in\mathbb M^2$ of Maya diagrams of zero total charge; and also (iii) by pairs $\lb \mathsf Y^+,\mathsf Y^-\rb\in\mathbb Y^2$ of Young diagrams corresponding to $\mathsf m^{+}$ and $\mathsf m^{-}$, and an integer $Q\equiv Q\lb\mathsf m^+\rb$. 
 
  The individual contributions can be readily computed using the Cauchy matrix representations (\ref{cauchyad}).
  \begin{theo}\label{thMaya}
  The Pain\-levé~III ($D_8$) tau function $\tau_{\mathrm{III}}\lb t\rb=\tau_{\mathrm{III}}\lb t\,|\,\sigma,\eta\rb$ from Theorem~\ref{theoFr} admits the following series representation:
  \beq
  \tau_{\mathrm{III}}\lb t\rb= 
  \sum_{\lb \mathsf p, \mathsf h\rb \,:\, \sharp\lb \mathsf p\rb=\sharp\lb \mathsf h\rb}
  e^{-4\pi i \eta Q}\;\Xi_{\mathsf p,\mathsf h}\lb\nu\rb
  \Delta^2_{\mathsf p,\mathsf h}\lb\nu\rb
  t^{\lb\nu-Q\rb^2+\left|\mathsf Y^+\right|+\left|\mathsf Y^-\right|},
  \eeq
  where $|\mathsf Y|$ denotes the total number of boxes in $\mathsf Y\in \mathbb Y$ and
  \begin{subequations}
  \begin{align}
  \label{Xiph}
  \Xi_{\mathsf p,\mathsf h}\lb\nu\rb=&\,
  \lb-1\rb^{Q}
  \frac{\Gamma^{2Q}\lb1+2\nu\rb}{
  \Gamma^{2Q}\lb1-2\nu\rb}
   \left[\prod_{\lb p,s'\rb\in\mathsf p}\lb p-\text{\footnotesize$\frac12$}\rb! \lb
    1-2s'\nu\rb_{p-\frac12}
    \prod_{\lb -q,s\rb\in\mathsf h}\lb q-\text{\footnotesize$\frac12$}\rb! \lb
      2s\nu\rb_{q+\frac12}\right]^{-2},\\
      \label{Deltaph}
  \Delta_{\mathsf p,\mathsf h}\lb\nu\rb=&\,\frac{\prod\limits_{ 
  \lb p,s'\rb<\lb \bar p,\bar s'\rb\in\mathsf p}
         \lb x_{p;s'}-x_{\bar p;\bar s'}\rb
   \prod\limits_{ \lb -q,s\rb<\lb -\bar q,\bar s\rb\in\mathsf h}
            \lb x_{-\bar q;\bar s}-x_{-q;s}\rb      }{
        \prod\limits_{\lb p;s'\rb\in\mathsf p}\prod\limits_{\lb -q;s\rb\in\mathsf h}
        \lb x_{p;s'}-x_{-q;s}\rb}.
  \end{align}
  \end{subequations}
  \end{theo}
  \pf From (\ref{cauchyad}) it follows that 
  \beq\label{auxcau01}
  \begin{aligned}
  \lb-1\rb^{\sharp\lb\mathsf p\rb}\operatorname{det}\mathsf a^{\,\mathsf p}_{\,\mathsf h}\operatorname{det}\mathsf d^{\;\mathsf h}_{\;
    \mathsf p}=&\,\lb-1\rb^{\sharp\lb\mathsf p\rb}\prod_{\lb p,s'\rb\in\mathsf p}\psi^{p;s'}\lb\nu\rb
    \bar\psi_{p;s'}\lb-\nu\rb \prod_{\lb -q,s\rb\in\mathsf h}
    \psi^{q;s}\lb -\nu\rb
        \bar\psi_{q;s}\lb \nu\rb 
      \times\\
      \times&\, \lb t^{\nu}e^{2\pi i \eta}\rb^{\sum_{\lb -q,s\rb\in\mathsf h} s-\sum_{\lb p,s'\rb\in\mathsf p}s'}
      t^{\sum_{\lb -q,s\rb\in\mathsf h} q+\sum_{\lb p,s'\rb\in\mathsf p}p}  \Delta^2_{\mathsf p,\mathsf h}\lb\nu\rb.
  \end{aligned}
  \eeq
  The power of $t^{\nu}e^{2\pi i \eta}$ can be further transformed as
  \ben
  \sum_{\lb -q,s\rb\in\mathsf h} s-\sum_{\lb p,s'\rb\in\mathsf p}s'=\sharp\lb \mathsf h^+\rb-\sharp\lb \mathsf h^-\rb-\sharp\lb \mathsf p^+\rb+\sharp\lb \mathsf p^-\rb=Q\lb \mathsf m^-\rb-
  Q\lb \mathsf m^+\rb=-2Q.
  \ebn
  It may be also easily shown (see Fig. 13 in \cite{GL16}) that
  \ben
  \sum\limits_{\lb -q,s\rb\in\mathsf h^\pm} q+\sum\limits_{\lb p,s'\rb\in\mathsf p^\pm}p
  =\frac{Q^2}{2}+\left|\mathsf Y^\pm\right|,
  \ebn
  so that the power of $t$ in the second line of (\ref{auxcau01}) becomes $Q^2+\left|\mathsf Y^+\right|+\left|\mathsf Y^-\right|$.
  The prefactor $\Xi_{\mathsf p,\mathsf h}\lb\nu\rb$ is obtained from the diagonal products in the first line by simple algebra.
  \epf
  
  \subsection{Nekrasov functions}
  In this subsection we rewrite the factorized expressions $\Xi_{\mathsf p,\mathsf h}\lb\nu\rb
    \Delta^2_{\mathsf p,\mathsf h}\lb\nu\rb$ 
   for the coefficients of the tau function expansion in a notation close to gauge theory. The main tool we need is a technical statement that can be found, for example, in \cite{GL16,GMfer}. In order to formulate it, let $\lb \mathsf Y^s,Q^s\rb\in \mathbb Y\times\Zb$ (with $s=\pm$)  be two charged Young diagrams, not necessarily the same as above. Denote by $\mathsf m^{s}\in \mathbb M$ the associated Maya diagrams. At this point we do not need to assume that $Q^++Q^-=0$. 
   
\begin{figure}[h!]
          \centering
          \includegraphics[height=3cm]{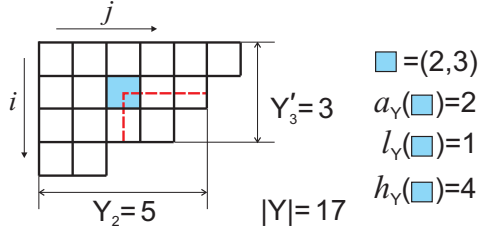}
          \begin{minipage}{0.8\textwidth}
          \caption{\label{piiiYD}
          Characteristics of Young diagrams.}
         \end{minipage}
          \end{figure}

   Introduce the following three quantities:
   \begin{enumerate}
   \item An explicit factorized function
\begin{gather}
  \label{apeq2}
  \begin{aligned}
  \tilde Z_{\,\mathsf{bif}}\lb\nu\,\bigl|\,\mathsf Y^+,Q^+;\mathsf Y^-,Q^-\rb=&\,\prod_{-q\in \mathsf h^+}\lb-\nu\rb_{q+\frac12}
  \prod_{-q\in\mathsf h^-}
  \lb\nu+1\rb_{q-\frac12}
  \prod_{p\in \mathsf p^-}\lb -\nu\rb_{p+\frac12}
  \prod_{p\in\mathsf p^+}\lb\nu+1\rb_{p-\frac12}\times \\
  \times &\,
  \frac{\;\prod\limits_{-q\in\mathsf h^+}\prod\limits_{p\in\mathsf p^-}\lb\nu-q-p\rb\prod\limits_{-q\in\mathsf h^-}\prod\limits_{p\in\mathsf p^+}\lb\nu+p+q\rb}{
  \prod\limits_{-q'\in\mathsf h^-}\prod\limits_{-q\in\mathsf h^+}\lb\nu-q+q'\rb\prod\limits_{p'\in \mathsf p^-}\prod\limits_{p\in\mathsf p^+}\lb\nu+p-p'\rb\;\;}.
  \end{aligned}
  \end{gather}
   which, as we will see in a moment, constitutes the main building block of $\Xi_{\mathsf p,\mathsf h}\lb\nu\rb
     \Delta^2_{\mathsf p,\mathsf h}\lb\nu\rb$.
  \item Another factorized expression, representing the Nekrasov bifundamental contribution:   
    \beq\label{ZNbif}
    Z_{\,\mathsf{bif}}\lb\nu\,|\,\mathsf Y^+,\mathsf Y^-\rb:=\prod_{\square\in \mathsf Y^+}\bigl(\nu+1+a_{\mathsf Y^+}\lb\square\rb+l_{\mathsf Y^-}\lb\square\rb\bigr)
    \prod_{\square\in \mathsf Y^-}\bigl(\nu-1-a_{\mathsf Y^-}\lb\square\rb-l_{\mathsf Y^+}\lb\square\rb\bigr),
    \eeq
    where $\mathsf Y^{\pm}\in\mathbb Y$ and the notation for Young diagrams follows Fig.~\ref{piiiYD}. The expressions $a_{\mathsf Y}\lb \square\rb$, $l_{\mathsf Y}\lb \square\rb$ and $h_{\mathsf Y}\lb \square\rb$ represent the arm-, leg-, and hook length of the box $\square $ in $\mathsf Y\in\mathbb Y$. In the case where $\square=\lb i,j\rb$ does not belong to $\mathsf Y$, the definition of the former two quantities is extended by $a_{\mathsf Y}\lb \square\rb=\mathsf Y_i-j$ and $l_{\mathsf Y}\lb \square\rb=\mathsf Y'_j-i$. In particular, we have
    \beq\label{zidents}
    \begin{aligned}
    Z_{\,\mathsf{bif}}\lb-\nu\,|\,\mathsf Y^-,\mathsf Y^+\rb=&\,(-1)^{|\mathsf Y^+|+|\mathsf Y^-|}Z_{\,\mathsf{bif}}\lb\nu\,|\,\mathsf Y^+,\mathsf Y^-\rb,\\
    Z_{\,\mathsf{bif}}\lb 0\,|\,\mathsf Y,\mathsf Y\rb=&\,(-1)^{|\mathsf Y|}\prod_{\square\in \mathsf Y}h^2_{\mathsf Y}\lb \square\rb.
    \end{aligned}
    \eeq
   \item For $Q\in\Zb$, define the ``structure constant'' $\Upsilon\lb \nu\,|\,Q\rb$  by
     \eq{
  \Upsilon\lb \nu\,|\,Q\rb=\frac{\Gamma^Q\lb 1+\nu\rb G\lb 1+\nu\rb}{G\lb 1+\nu+Q\rb}.
   \label{constC}
     }
     Here $G\lb z\rb$ denotes the Barnes $G$-function satisfying the relation $G\lb z+1\rb=\Gamma\lb z\rb G\lb z\rb$. Note that $\Upsilon\lb \nu\,|\,Q\rb$ is actually a rational function of $\nu$.
   \end{enumerate}
   \begin{lemma}\label{lemma42}
   We have  
   \beq\tilde Z_{\,\mathsf{bif}}\lb\nu\bigl|\,\mathsf Y^+,Q^+;\mathsf Y^-,Q^-\rb=\pm \Upsilon^{-1}\lb\nu\,\bigl|\,Q^+-Q^-\rb
     Z_{\,\mathsf{bif}}\lb\nu+Q^+-Q^-\,\bigl|\,\mathsf Y^+,\mathsf Y^-\rb,
   \label{id2}
   \eeq
   where $\pm$ means that the equality holds up to an overall sign.
   \end{lemma}
   \pf See \cite[Appendix A]{GL16}. \epf
   
  Lemma~\ref{lemma42} thus relates certain products over boxes of Young diagrams to some explicit functions
  of particle and hole coordinates in the relevant Maya diagrams. Let us now use it to identify the  corresponding Nekrasov functions in (\ref{thMaya}). We first prove 
 \begin{lemma}
 We have 
\beq
\Xi_{\mathsf p,\mathsf h}\lb\nu\rb  \Delta^2_{\mathsf p,\mathsf h}\lb\nu\rb=\frac{\Gamma^{2Q}(1+2\nu)}{\Gamma^{2Q}(1-2\nu)}
\frac{\Upsilon\lb 2\nu\,|\,-2Q\rb \Upsilon\lb -2\nu\,|\,2Q\rb}{\prod\limits_{s,s'=\pm1}Z_{\,\mathsf{bif}}\lb(Q-\nu)(s'-s)\,|
\,\mathsf Y^{s'},\mathsf Y^s\rb}.
\label{id3}
\eeq   
 \end{lemma}
 \pf
First of all, one may further decompose $\Delta_{\mathsf p,\mathsf h}$:
\ben
\Delta_{\mathsf p,\mathsf h}(\nu)=\Delta^{++}_{\mathsf p,\mathsf h}\Delta^{+-}_{\mathsf p,\mathsf h}(\nu)\Delta^{--}_{\mathsf p,\mathsf h},
\ebn
where
\begin{align*}
  \Delta^{\pm\pm}_{\mathsf p,\mathsf h}&=\,\frac{\prod\limits_{ p,\bar p\in\mathsf p^{\pm}: p<\bar p}
         \lb p-\bar p\rb
   \prod\limits_{-q,-\bar q\in\mathsf h^{\pm}:q>\bar q}
            \lb q-\bar q\rb      }{
        \prod\limits_{p\in\mathsf p^{\pm}}\prod\limits_{ -q\in\mathsf h^{\pm}}
        \lb p+q\rb},
\\
  \Delta^{+-}_{\mathsf p,\mathsf h}\lb\nu\rb&=\,\frac{\prod\limits_{  p_+\in\mathsf p^+}\prod\limits_{ p_{-}\in\mathsf p^-}
         \lb -2\nu+p_+-p_-\rb
   \prod\limits_{  -q_+\in\mathsf h^+}
   \prod\limits_{   -q_-\in\mathsf h^-}
            \lb -2\nu -q_++q_-\rb      }{
        \prod\limits_{p_+\in\mathsf p^+}\prod\limits_{ -q_-\in\mathsf h^-}
        \lb -2\nu+p_++q_-\rb\;\;
\prod\limits_{ p_-\in\mathsf p^-}\prod\limits_{-q_+\in\mathsf h^+}
        \lb 2\nu+p_-+q_+\rb\;\;\;}.
\end{align*}
Comparing   (\ref{apeq2}) with (\ref{Deltaph}), we can write
\eq{
\begin{split}
&\prod\limits_{s,s'=\pm1}\tilde Z_{\,\mathsf{bif}}\lb\nu \lb s- s'\rb\,|\,\mathsf Y^{s'},Qs';\mathsf Y^s,Qs\rb=\pm\left[\Delta^{++}_{\mathsf p,\mathsf h}
\Delta^{+-}_{\mathsf p,\mathsf h}(\nu)\Delta^{--}_{\mathsf p,\mathsf h}\right]^{-2}\times\\&\times
  \prod_{-q\in\mathsf h^-}\lb-2\nu\rb_{q+\frac12}
  \prod_{-q\in\mathsf h^+}\lb2\nu+1\rb_{q-\frac12}
  \prod_{p\in\mathsf p^+}\lb -2\nu\rb_{p+\frac12}
  \prod_{p\in\mathsf p^-}\lb2\nu+1\rb_{p-\frac12}\times\\&\times
  \prod_{-q\in\mathsf h^+}\lb2\nu\rb_{q+\frac12}
  \prod_{-q\in\mathsf h^-}\lb-2\nu+1\rb_{q-\frac12}
  \prod_{p\in\mathsf p^+}\lb 2\nu\rb_{p+\frac12}
  \prod_{p\in\mathsf p^+}\lb-2\nu+1\rb_{p-\frac12}\times\\&\times
  \left[\prod_{-q\in\mathsf h^-}\lb q-\text{\footnotesize $\frac12$}\rb!\prod_{-q\in\mathsf h^+}\lb q-\text{\footnotesize $\frac12$}\rb!  \prod_{p\in\mathsf p^-}\lb p-\text{\footnotesize $\frac12$}\rb!\prod_{p\in\mathsf p^+}\lb
  p-\text{\footnotesize $\frac12$}\rb !\right]^2.
\label{id1}
\end{split}
}
Using the identity $(z)_{q+\frac12}=z\cdot(z+1)_{q-\frac12}$ for the Pochhammer's symbol, the balance condition
$\sharp\lb \mathsf h^+\rb+\sharp\lb \mathsf h^-\rb=\sharp\lb \mathsf p^+\rb+\sharp\lb \mathsf p^-\rb$, and comparing the last three lines with (\ref{Xiph}), we
can rewrite (\ref{id1}) as
\eq{
\prod\limits_{s,s'=\pm1}\tilde Z_{\,\mathsf{bif}}\lb\nu \lb s-s'\rb\,|\,\mathsf Y^{s'},Qs';\mathsf Y^s,Qs\rb=\pm\frac{\Gamma^{2Q}(1+2\nu)}{\Gamma^{2Q}(1-2\nu)}\,
\Xi^{-1}_{\mathsf p,\mathsf h}\lb\nu\rb  \Delta^{-2}_{\mathsf p,\mathsf h}\lb\nu\rb.
}
Combining this result with (\ref{id2}), we immediately obtain (\ref{id3}) up to an overall sign. It suffices to check it for real $\nu\notin\Zb$. For the left side this sign is obviously equal to $\lb -1\rb^Q$.
From the identities (\ref{zidents}) it follows that the right side of (\ref{id3}) may be rewritten as
\ben
\frac{G\lb 1+2\nu\rb G\lb 1-2\nu\rb}{G\lb 1+2\nu-2Q\rb G\lb
1-2\nu+2Q\rb}
\left[Z_{\,\mathsf{bif}}\lb 2\nu-2Q\,|\,\mathsf Y^-,\mathsf Y^+\rb \prod_{\square\in \mathsf Y^+}h_{\mathsf Y^+}\lb \square\rb \prod_{\square\in \mathsf Y^-}h_{\mathsf Y^-}\lb \square\rb\right]^{-2}.
\ebn
Its sign is therefore determined by the Barnes function prefactor in the last expression, and can be easily shown to be $\lb -1\rb^Q$.
\epf

We can now formulate our final result.
\begin{theo}\label{thYoung}
Let $\mc Z_{\mathrm{SU}\lb 2\rb}\lb t\,|\,\nu\rb$ be the Nekrasov instanton partition function of the pure gauge theory, defined as a double sum over partitions:
\beq
\mc Z_{\mathrm{SU}\lb 2\rb}\lb t\,|\,\nu\rb=C\lb\nu\rb\sum_{\mathsf Y^+,\mathsf Y^-\in\mathbb Y}
\frac{t^{\nu^2+|\mathsf Y^+|+|\mathsf Y^-|}}{\prod\nolimits_{s,s'=\pm1}Z_{\,\mathsf{bif}}\lb\nu(s-s')\,|
\,\mathsf Y^{s'},\mathsf Y^s\rb},
\eeq
with $C\lb\nu\rb= \left[\prod\nolimits_{s=\pm1}G\lb 1+2s\nu\rb\right]^{-1}$ and $Z_{\,\mathsf{bif}}\lb\nu\,|\,\mathsf Y',\mathsf Y\rb$ defined by (\ref{ZNbif}). The dual partition function
\beq
\mc Z^{\mathrm{dual}}_{\mathrm{SU}\lb 2\rb}\lb t\,|\,\nu,\eta\rb=\sum_{n\in\Zb}e^{4\pi i n\eta}\mc Z_{\mathrm{SU}\lb 2\rb}\lb t\,|\,\nu+n\rb
\eeq
admits Fredholm determinant representation
\beq
\mc Z^{\mathrm{dual}}_{\mathrm{SU}\lb 2\rb}\lb t\,|\,\nu,\eta\rb=C\lb\nu\rb t^{\nu^2}\operatorname{det}\lb \mathbb 1-K\rb,
\eeq
where $K$ is the generalized Bessel kernel from Theorem~\ref{theoFr} (with $\sigma=\nu-\frac12$), and thereby coincides with the general tau function of the Painlevé III ($D_8$) equation.
\end{theo}

 \end{document}